\documentclass[preprint,12pt,3p]{elsarticle}

\usepackage{matlab-prettifier}
\usepackage{xfrac}
\usepackage{graphicx}
\usepackage{parskip}
\usepackage{hyperref}
\usepackage{longtable}
\usepackage{adjustbox}
\usepackage{setspace}
\hypersetup{ 
backref=true, % link backreferences (i.e. citations to their list)
bookmarks=true, % show bookmarks bar?
unicode=false, % non-Latin characters in Acrobats bookmarks
pdftoolbar=true, % show Acrobats toolbar?
pdfmenubar=true, % show Acrobats menu?
pdffitwindow=false, % window fit to page when opened
pdfstartview={FitH}, % fits the width of the page to the window
pdftitle={}, % title
pdfauthor={}, % author
pdfsubject={}, % subject of the document
pdfkeywords={}{}, % list of keywords
pdfnewwindow=true, % links in new window
colorlinks=true, % false: boxed links; true: colored links
linkcolor=BlueViolet, % colour of internal links
citecolor=maroon, % colour of links to bibliography
urlcolor=blue,
}
\usepackage{comment}
\usepackage{subfig}
\usepackage{caption}

\usepackage{amssymb}
\usepackage{gensymb}

\usepackage{natbib}
\biboptions{comma,round}
\setcitestyle{authoryear}

\begin{document}
%\begin{spacing}{1.5}
\begin{frontmatter}

\title{Navigating the energy trilemma during geopolitical and environmental crises\footnote{Dina Azhgaliyeva and Zhanna Kapsalyamova had constructive comments on a previous version. The Asian Development Bank provided welcome financial support.}}

\author[label1,label2,label3,label4,label5,label6]{Richard S.J. Tol}
\address[label1]{Department of Economics, Jubilee Building, University of Sussex, Falmer, BN1 9SL, United Kingdom; r.tol@sussex.ac.uk}
\address[label2]{Institute for Environmental Studies, Vrije Universiteit, Amsterdam, The Netherlands}
\address[label3]{Department of Spatial Economics, Vrije Universiteit, Amsterdam, The Netherlands}
\address[label4]{Tinbergen Institute, Amsterdam, The Netherlands}
\address[label5]{CESifo, Munich, Germany}
\address[label6]{Payne Institute for Earth Resources, Colorado School of Mines, Golden, CO, USA}

\begin{abstract}
There are many indicators of energy security. Few measure what really matters\textemdash affordable and reliable energy supply\textemdash and the trade-offs between the two. Reliability is physical, affordability is economic. Russia's latest invasion of Ukraine highlights some of the problems with energy security, from long-term contracts being broken to supposedly secure supplies being diverted to retired power plants being recommissioned to spillovers to other markets. The transition to carbon-free energy poses new challenges for energy security, from a shift in dependence from some resources (coal, oil, gas) to others (rare earths, wind, sunshine) to substantial redundancies in the energy capital stock to undercapitalized energy companies, while regulatory uncertainty deters investment. Renewables improve energy security in one dimension, but worsen it in others, particularly long spells of little wind. Security problems with rare earths and borrowed capital are less pronounced, as stock rather than flow.

\end{abstract}

\end{frontmatter}

\section{Introduction}
\label{sc:intro}
We want energy to be cheap, reliable, and clean. It is typically easy to meet one of these three criteria, but meeting all three at the same time is difficult. This is known as the \emph{energy trilemma}: You can't have your cake, eat it, and consume it. Trade-offs are real.

The energy trilemma and its components are not new. As the price of fuel wood rose in London, people switched to sea coal, bituminous coal mined on the northeast coast of England. Burning this made the air intolerable to breathe, and in 1307 King Edward I of England banned the use of sea coal in lime kilns. The ban was no success and later kings and parliaments issued their own regulations \citep{teBrake1975}. Nonetheless, in December 1952, some 4,000 people were killed by air pollution and maybe 8,000 more in the following months \citep{Bell2001}. The Clean Air Act of 1956 marks the beginning of the transition away from coal as the prime fuel for heating in the cities of the UK. Elsewhere, indoor and outdoor air pollution, primarily due to energy use, continue to kill millions of people each year.\footnote{See \href{https://www.who.int/health-topics/air-pollution\#tab=tab_2}{WHO}.}

\emph{Energy security} focuses on two aspects of the energy trilemma: reliability and affordability \citep{Lefevre2010}. The two concepts are often mixed together,\footnote{Some analysts define the energy trilemma to be a three-way trade-off between cleanliness, security and equity. Besides combining reliability and affordability, equitable access to energy is a matter of the distribution of income\textemdash unless there is strong price discrimination. Others replace clean energy with sustainable energy, but why use a long word where a short one will do? Besides, sustainability nowadays no longer refers just to the environment but also includes notions of social justice and economic development \citep{Purvis2019}.} but they are really separate. Energy reliability is a physical concept. Electric power is unreliable if transmission lines frequently fail, or generation plants suffer many outages. Reliable electric power is available when it is needed, unreliable electricity may or may not be there.

Affordability is an economic concept. Electricity may be available but sold at such a high price that its use is forgone or rationed. From the perspective of the final user, it does not matter whether energy is not there or not affordable. In either case, energy is not used. From an analytical perspective, however, it is important to distinguish the two concepts because technical solutions differ and the two objectives may clash. The reliability of an electricity grid can be improved with more transmission lines, redundant except in emergencies. The electricity supply can be made more reliable by adding more power plants, used only in times of exceptional outages or very high demand. Such an increase in reliability would come at a cost, and so make electricity less affordable.

\citet{Jansen2010} argue that environmental externalities pose security risks, so that energy security encompasses the whole of the energy trilemma. \citet{Bohi1993, Bohi1996} agree but restrict attention to those externalities that can be meaningfully influenced by policy, as indicators that are impervious to policy intervention should not be used to advise policy. I disagree, not because clean energy is not important, but because indicators should support policy by clarifying choices and consequences. A single indicator that obscures the reality of the energy trilemma is not helpful. Free after \citet{Tinbergen1952}, we should have as many indicators as we have problems.

This paper continues as follows. Section \ref{sc:indices} discusses the various indicators of energy security that have been used in the literature. Section \ref{sc:instruments} treats policy instruments to increase energy security. Section \ref{sc:war} is about the effects of geopolitics on energy reliability and affordability. Section \ref{sc:climate} treats the impact of climate policy in the short- and long-run. Section \ref{sc:conclude} concludes.

\section{Indicators of energy security}
\label{sc:indices}
If we want to assess the impacts of geopolitics and climate policy on energy security, we need to be able to measure energy security. Much thought has gone into this, but this has not brought much clarity. 

\citet{Bohringer2015} critically reviews energy security indicators and their use in policy. Energy security indicators tend to suffer from the following limitations. First, indicators are supply-oriented, disregard the demand side \citep{Jansen2010, Sovacool2013, Gracceva2014}. Second, indicators are proxies only, do not assess energy system's responses to shocks \citep{Cherp2011, Gracceva2014}. Third, energy security indicators have no information on the costs and benefits of different levels of energy security \citep{Gracceva2014}. Fourth, different energy security indicators cannot meaningfully be added or compared \citep{Bohringer2007, Kruyt2009, Frondel2014}.

According to \citet[][see also \citet{Kruyt2009, Loschel2010}]{Bohringer2015}, four indicators of energy security are in widespread use:
\begin{itemize}
    \item \textbf{Primary energy intensity} is defined as total primary energy use over Gross Domestic Product (GDP). Primary energy use is a physical measure, so this indicator only proxies reliability, not affordability. However, this indicator makes no distinction between more reliable and less reliable energy supplies. No account is taken of international trade in energy; offshoring energy-intensive industry would seem to increase energy security, even if goods are now imported from countries with a higher energy intensity \citep{Gnansounou2008}. GDP too is problematic. It can be a poor measure of economic output in small open economies. Comparison of prices across international borders is difficult too; economies vary greatly in size between market exchange rates and purchasing power ones \citep{Samuelson2014, Suehiro2008}. 
    \item \textbf{Dependence on foreign primary energy supply} is defined as the sum (over all fuels) of net imports (or zero for net exporters) divided by total primary energy use \citep{Bhattacharyya2011, LeCoq2009}. As with the previous indicator, there is no distinction between more and less reliable energy supplies. All foreign suppliers are deemed to be equally risky, and all domestic suppliers are supposed to be without risk.
    \item \textbf{Concentration of primary energy supply} is defined as the Herfindahl-Hirschman index\textemdash the sum of squared market shares\textemdash for fuels \citep{Bhattacharyya2011}. This indicator again ignores the actual reliability of the energy supply. It also ignores that different fuels serve different purposes\textemdash liquid fuels for transport, solid and gaseous fuels for power generation\textemdash so that differences in demand naturally lead to more or less reliance on particular fuels \citep{Stirling2010}.
    \item \textbf{Concentration of foreign primary energy supply} is defined as the Herfindahl-Hirschman index for net energy imports, where concentration is measured either over the number of foreign suppliers \citep{Kleindorfer2005} or over the number of foreign suppliers and fuels \citep{Frondel2014}. Once more, reliability of imports is omitted from the indicator\textemdash oil purchases from Norway are treated the same as oil purchases from Libya\textemdash although this can be accommodated by introducing a riskiness parameter per supplier. Fungibility is another critique \citep{LeCoq2009}. A country may depend on a single supplier of coal. An ideosyncratic shock to that supplier would not be a problem if other suppliers can take over. A country may buy oil from many suppliers, but this would not protect it from a system-wide shock. Transport is ignored too. Crossing the territory of a third party may be risky\textemdash recall piracy off the Horn of Africa and hijackings in the Strait of Hormuz\textemdash and some modes of transport are more flexible than others\textemdash contrast gas pipelines and liquified natural gas.
\end{itemize}

\citet{Ang2015} also review the literature on energy security, finding no fewer than 83 different definitions and a great many indicators. For instance, \citet{Sovacool2011} use 320 simple indicators and 52 complex ones. None of this makes much sense. Energy security is security from a human perspective, and it should therefore be measured as a reduction of human welfare \citep{Bohi1996}. There are two reasons why energy might not be secure: an energy source is temporarily or permanently (\textit{i}) unavailable or (\textit{ii}) unaffordable, and cannot be replaced at short notice.

\textit{Ex post}, energy security is easy to observe: a power plant tripped, an oil tanker ran aground, energy bills were unpaid, energy offers not bought. \textit{Ex ante}, energy reliability is hard to measure because it necessarily involves an assessment of the probability of things not going as expected. Energy affordability is predictable to the extent that incomes and energy prices are.

As argued above, energy reliability is a physical concept: Is the primary energy available, can it be transformed to a useful energy carrier, and can it be transported to the final user? Energy affordability is an economic concept: Is the price acceptable to the final user?

\subsection{\textit{Ex post} indicators}
The World Bank has a number of indicators that measure the reliability of the electricity supply, including:
\begin{itemize}
    \item fraction of value lost due to electrical outages;
    \item percentage of firms experiencing electrical outages; and
    \item monthly number of power outages in firms.
\end{itemize}
Electricity is important but there are other energy sources as well. It would be good if someone would systematically collect information on shortages of transport, heating, and cooking fuels.

The negative impact of power outages is well-documented for both firms \citep{Pasha1989, Beenstock1991, Tishler1993, Beenstock1997, Serra1997, Steinbuks2010, Alby2013, Allcott2016, Cole2018, Elliott2021, Chen2022} and households \citep{Carlsson2007, Carlsson2008, Carlsson2011, Amador2013, Chakravorty2014, Ozbafli2016, Poczter2017, Kennedy2019, Meles2020, Bajo-Buenestado2021, Carlsson2021, Deutschmann2021, Meles2021, Motz2021, Sedai2021, Alberini2022, Aweke2022, Lawson2022, Toto2022}, as well as for the economy as a whole \citep{Sanghvi1982, deNooij2007, deNooij2009, Andersen2013, Reichl2013, Carranza2021, Woo2021}. The evidence is for all parts of the world, and all levels of development. The negative impact comes in two parts. Unreliable electricity leads to interruption of production and daily life. In addition, firms invest in expensive back-up equipment, locking scarce capital into unproductive means.

These studies make clear that an unreliable energy supply is bad for the economy in the short-run and for economic development in the long-run\textemdash capital diverted to back-up power generation could have been used more productively; learning and human capital accumulation are interrupted too. Although the actual outages are by chance, the probability can be reduced by better management of power plants and transmission lines and by better regulation of utilities.

The World Bank also publishes data on access to energy:
\begin{itemize}
     \item Fraction of people (total, rural, urban) that have access to electricity; and
     \item Fraction of people (total, rural, urban) that have access to clean energy.
\end{itemize}
It would be good to extend this to access to modern energy. Furthermore, while energy access is an important issue in poorer countries, energy poverty is important in richer countries\textemdash but there is no systematic data collection on that. Energy poverty is variously defined as energy expenditures above a certain fraction of income or an inability to provide a basic level of energy services. The latter, better definition is difficult to measure consistently over time and space.\footnote{Energy poverty is reported in poorer countries as well \citep{Barnes2011, Khandker2012, Andadari2014, Sadath2017, Crentsil2019, Feeny2021, Gafa2021} but difficult to separate from energy access.}

The impact of energy access is well-documented too, with mostly positive effects on a range on economic, social and environmental aspects across the world, for all levels of development, and in the short- and long-term \citep{Dinkelman2011, Grogan2013, Lipscomb2013, Rao2013, Khandker2014, Dasso2015, Grimm2015, Kitchens2015, Grogan2016, Peters2016, Salmon2016, Abeberese2017, Akpandjar2017, Barron2017, DaSilveiraBezerra2017, Grimm2017, vandeWalle2017, Aklin2018, Burke2018, Ding2018, Fujii2018, Grogan2018, Kumar2018, Lewis2018, Rathi2018, Saing2018, Thomas2018, Dang2019, He2019, JahangirAlam2019, Litzow2019, Richmond2019, Zhang2019, Burgess2020, Cravioto2020, Diallo2020, Emmanuel2020, Fujii2020, Irwin2020, Lee2020, Lee2020JEP, Lewis2020, Sievert2020, Tagliapietra2020, Thomas2020, Acheampong2021, Acheampong2021EE, Chhay2021, Fried2021, Gaggl2021, Gupta2021, Jeuland2021, Sedai2021EE, Wagner2021, Wirawan2021, Wu2021, Acharya2022, Adom2022, Ayana2022, Bo2022, Chaurey2022, Dendup2022, Hong2022, Koirala2022, Ogunro2022, Sedai2022}.

These papers show that improving access to energy, by expanding the physical supply and reducing prices, has a direct effect on the economy by reducing costs, freeing up time, and facilitating more production. In addition, it improves health care and education which, in the long term, further accelerate economic development.

Energy poverty too has negative impacts on well-being \citep{Sambodo2019, AwaworyiChurchill2020welfare, Nie2021}, physical health \citep{Teller-Elsberg2016, Ortiz2019, Bukari2021, AwaworyiChurchill2021, Nawaz2021, Prakash2021}, mental health \citep{Zhang2021}, education \citep{Oum2019, Rafi2021, Apergis2022}, crime \citep{Hailemariam2021, AwaworyiChurchill2022crime}, agriculture \citep{Shi2022}, and development in general \citep{Singh2020, Acheampong2021}.

\section{Improving energy security}
\label{sc:instruments}
Policymakers have a number of instruments at their disposal to improve energy security in its many guises. This is not the place for an exhaustive discussion  \citep[see, for instance,][]{Anderson2019, Baumol1988, Berck2011, Endres2012, Field2009, Goodstein2005, Hanley2007, Hanley2013, Harris2018, Hodge1995, Kahn2020, Keohane2016, Kolstad2011, Lewis2019, Pearce1990, Perman2011, Phaneuf2017, Tietenberg2018, Turner1994, Wills1997}.

The reliability of the energy supply is threatened in two ways: There may not be enough energy, or the energy may not reach its destination.

Insufficient capacity is a particular problem in power generation. The technical answer is more capacity. As electricity cannot (yet) be stored at scale and demand varies considerably during the day, week and year, there is typically a mismatch between peak demand and maximum supply. Peak demand lasts only a few hours, a short period to earn back the investment in peak supply. The best solution is a mixture of setting a level of acceptable blackouts and preparing for those and a reverse auction to buy spare generating capacity, financed by a levy on electricity use \citep{Creti2007}.

Inadequate transport or transmission is the other cause of an unreliable energy supply. The technical solution is to build redundancy in transport and transmission systems. If the market is competitive\textemdash e.g., tanker transport of oil\textemdash the costs of redundancy will be weighed against the costs of non-delivery, a breach of contract, and loss of reputation. If the market is not competitive\textemdash power cables, pipelines\textemdash direct regulation is the way forward. Natural monopolies tend to be state-owned and strictly regulated anyway, so additional regulation is straightforward whereas price signals\textemdash taxes or subsidies\textemdash are less effective without competition \citep{Jamasb2012, Schmidthaler2015}. 

Energy affordability is about energy access in poorer countries and about energy poverty in richer ones. In both cases, poverty is the core problem. Rich people in poor countries have access to modern fuels. Energy companies happily hook up neighbourhoods once enough people can pay for their products. Similarly, energy poverty in rich countries is tightly correlated with income poverty. Stimulating economic growth, and particularly economic growth that disproportionally favours the less well-off is therefore a key strategy to improve energy affordability.

More targeted interventions are also possible. Many countries subsidies energy use. Price subsidies are not advised. Price subsidies help those who would otherwise not be able to afford energy, but also and primarily those who would have bought energy at the unsubsidized price anyway. Price subsidies also encourage waste when energy is, in fact, short. Price vouchers allow targeted price support \citep{Podesta2021}. Income support is another, better alternative to price support\textemdash if well targeted \citep{Best2021, GarciaAlvarez2021, Bagnoli2022}.

Furthermore, as lack of energy access and energy poverty hold back development (see above), alleviating this should be part of an overall economic development strategy \citep{Bouzarovski2012, Karpinska2021}. Unaffordable energy is often caused by a lack of investment, in turn caused by a lack of access to capital markets. Investment subsidies are thus justified, in home insulation and efficient heating in richer countries, and in microgeneration, -grids and -storage in poorer countries. However, energy poverty is not just about financials \citep{Xu2019, AwaworyiChurchill2020ethnic, Karpinska2020, Ampofo2021, Paudel2021, AwaworyiChurchill2022religion, Dogan2022, Koomson2022, Koomson2022ethnic, Moniche-Bermejo2022}. Any campaign against energy poverty needs to pay careful attention to age and family structure as well as to ethnic, racial, and religious discrimination.

\section{Geopolitics and energy security}
\label{sc:conflict}
The exploitation of fossil fuels, and particularly of oil and gas, is heavily concentrated in a small number of places. Although there are oil and gas fields in many countries, most are relatively small. A few large producers dominate production. This has been the case since the start of the large-scale use of oil and gas.

The concentration of production implies that political unrest or violent conflict at the locale of oil and gas fields has a disproportionate impact on the world market for oil and gas. The concentration of production increases the importance of long-distance transport and the bottlenecks of international trade, such as the Panama and Suez Canals, and the Straits of Hormuz and Malacca. Furthermore, aware of the strategic importance of the centres of oil and gas production, outside forces have long sought to control these centres, or control the strongmen that control them, competing with indigenous people and with other outsiders. In return, the strongmen have sought to influence the politics of other countries, both near their borders and far away. The result is a vicious circle of political instability, interspersed with periods of stable but brutal and brittle regimes.

The second invasion of Ukraine by Russia illustrates the short-term issues. The violence has affected key energy infrastructure\textemdash damage to substations and thermal generators; threats to nuclear plants and hydropower dams\textemdash some accidental, and some apparently deliberate. The violence is concentrated in Ukraine, but occasionally spills into Russia and there are seemingly related acts of sabotage in Germany and the Baltic Sea. The rulers of Russia may have hoped that its position as the main supplier of energy to Europe would prevent other countries to come to Ukraine's aid, but that was a miscalculation. The flow of oil and gas from Russia to Europe fell sharply. This forced the countries of Europe to seek imports from elsewhere, driving up the price of oil and liquefied natural gas (LNG). This in turn made energy unaffordable elsewhere. Pakistan, for instance, could no longer afford to import LNG and suffered power blackouts as a result. At the same time, Russian oil and gas traded at a discount, benefiting those countries that had the infrastructure to import (e.g., gas pipelines) and even re-export. The details are different for other conflicts, but violent conflict involving large energy exporters causes a lot of misery.

For the effects in the long-run, the literature on the natural resource curse offers some empirical support for the hypothesis that economies with weak institutions and an abundance of oil and gas, grow more slowly as they are susceptible to political corruption \citep{Ross1999, Sachs2001, Jensen2004, Papyrakis2004, Bulte2005, Hodler2006, Mehlum2006EJ, Mehlum2006, Robinson2006, Boschini2007, Brunnschweiler2008WD, Kolstad2009, vanderPloeg2009, Torvik2009, Aslaksen2010, Vicente2010, Cavalcanti2011, VanDerPloeg2011, Williams2011, Boschini2013, Brollo2013, Betz2015, Havranek2016, Badeeb2017} and violent conflict \citep{Grossman1999, Collier2005, Dunning2005, Basedau2009, Brunnschweiler2009}\textemdash although there are also papers highlighting flaws in the research \citep{Brunnschweiler2008JEEM, Alexeev2009, vanderPloeg2010, Haber2011, Smith2015}

Unlike outages caused by the technical failure of energy production and transport, or power generation and transmission, unrest and conflict are hard to predict. Instead of objective probabilities based on observed frequencies, we have subjective degrees of belief that, as autocratic regimes are rarely transparent, are based on incomplete knowledge and understanding \citep{Garcia2012}. Yet, as once again demonstrated by the second Russian invasion of Ukraine in February 2022, geopolitical risks can, when realized, cause great havoc on energy markets as the impacts are system-wide rather than location-specific.

Although some have argued that a shift away from fossil fuels would lead to a reduction in geopolitical energy risks \citep{Kemfert2019}, others point out that geopolitical risks would change rather than disappear \citep{Hache2018}\textemdash see below.

\section{Climate policy and transition risk}
\label{sc:climate}
The energy trilemma has that we want energy that is reliable, affordable, and clean. Reliability and affordability together constitute energy security. The drive for cleaner energy affects its reliability and affordability. In the current discourse, ``clean'' energy is seen as carbon-free energy. There are other, perhaps larger environmental problems due to energy use\textemdash such as indoor air pollution, outdoor air pollution, and acidification\textemdash but these primarily affect poorer countries and are not seen as global priorities.

The replacement of fossil fuels by renewable energy will, in the long run, lead to a more reliable energy supply. Whereas thermal power plants are large and therefore few, wind turbines and solar panels are small and therefore many. By the law of large numbers, a large number of small power sources is less vulnerable to outages\textemdash be it due to mechanical faults, natural disasters, or terrorist attacks\textemdash than a small number of large power sources. Maintenance too is less disruptive.

On the other hand, solar and wind power are not dispatchable; power generation happens when it does, rather than when it needs to happen. This is particularly a problem for wind power. There is no solar power at night, but this is no surprise and can be solved with short-term electricity storage, as demand drops rapidly mid-evening. Lulls in wind can last for weeks, well beyond storage capacity, and may coincide with high demand\textemdash in Western Europe, for instance, winter cold and low winds go hand in hand. 

Some argue that renewable energy is more secure because it is mostly generated in the home country rather than imported.\footnote{See \href{https://en.wikipedia.org/wiki/Energy_security}{Wikipedia}, \href{https://www.gov.uk/government/publications/british-energy-security-strategy/british-energy-security-strategy}{UK Government}, \href{https://www.energy.gov/articles/doe-releases-first-ever-comprehensive-strategy-secure-americas-clean-energy-supply-chain}{US Government}.} This argument is false. Foreign suppliers are not necessarily less reliable than domestic ones. The argument either rests on xenophobia or on the false belief of being in control of what is happening in your own country.

Others argue that renewable energy is not secure because it relies on rare earths\footnote{See \href{https://www.gov.uk/government/publications/uk-critical-mineral-strategy/resilience-for-the-future-the-uks-critical-minerals-strategy}{UK Government}, \href{https://www.energy.gov/bil/rare-earth-security-activities}{US Government}} and depends on foreign capital \citep{Nakatani2010}. These arguments affect the speed of expansion of renewables rather than their functioning once installed. Rare earths are essential for both generation and storage. Their spatial concentration is a reason for concern. However, existing solar panels will continue to operate if the supply of rare earths is interrupted\textemdash unlike thermal plants which cease to operate if their fuel runs out.

The same argument holds for capital. Renewable energy uses more capital per kilowatthour than fossil energy and is therefore more exposed to movements of the interest rate and to sanctions in the capital market. The argument holds for the financing of new renewables, and for the refinancing of existing renewables. It does not hold for their operation. Operators run their wind turbines and solar panels regardless of the interest rate. The same cannot be said of thermal plants, which cease operation if the wholesale electricity price does not cover the cost of fuel.

\citet{Hache2018} argues that patents are another bottleneck\textemdash a country or company may deny another country or company a license for the use of advanced technology. But, as with rare earths and investment, withholding patents would decelerate the expansion of renewables but not stop existing renewables. And, anyway, legal niceties such as respect for intellectual property rights rapidly go out of the window in case of conflict.

An expansion of nuclear power would also help to reduce climate change, but probably at the expense of affordability and reliability. Taking the costs of accident prevention and waste disposal into account, nuclear fission is not among the cheaper sources of electricity \citep{Ahearne2011}. Nuclear power plants are large;\footnote{Nuclear fusion plants would be larger still.} unscheduled outages therefore threaten a reliable power supply. The situation in France in 2022 is a reminder. While small modular reactors are all the rage at the moment, these are in fact only somewhat smaller than a typical gas-fired power plant. Nuclear power poses two unique challenges. An expansion of the nuclear power supply large enough to have a notable effect on greenhouse gas emissions and so climate change would require the building of nuclear power plants in currently unstable countries. The first challenge is that the people who run a nuclear power plant, know how to and have the material to build a dirty bomb. Secondly, as illustrated by the second Russian invasion of Ukraine, it is a really bad idea to situate a nuclear power plant in a war zone.

However, while climate policy may make energy more reliable in the long run, this is not necessarily the case in the short and medium term. Instability in the fossil-fuel producing regions is one concern, as the old regimes lose their power of patronage\textemdash and the restraint to hit the buyers of their energy.

Another concern is the scale of investment needed, particularly if the ambitious goals set out in the Paris Agreement are to be met.\footnote{See \href{https://www.iea.org/reports/world-energy-outlook-2021/mobilising-investment-and-finance}{IEA}.} A rapid expansion of renewable energy by investors with a limited budget may well lead to a lack of redundancy. Reliability would fall as a consequence.

A rapid transition to renewables risks stranding fossil fuel assets \citep{Davis2010, Tong2019}. One can see this as an increase in redundancy, as delays between deactivation and demolition can be long. The current energy crisis in Europe due to Russia's second invasion of Ukraine is indeed alleviated by previously mothballed power stations being turned back on. However, the greater risk to energy security is the higher probability of bankruptcy as companies have to retire assets before the end of their economic lifetime. This leaves less money to invest within the energy sector and deters money from outside the sector from flowing in. If governments bail out energy companies, the budget for energy support falls\textemdash including investment in such things as peak capacity, transmission, interconnection, and storage.

Besides the impact on the reliability of the energy supply, climate policy also affects the affordability of energy.\footnote{\citet{Chakravarty2013} note that lifting 3.5 billion people out of energy poverty would raise the global mean surface air temperature by 0.13\celsius{} only.} Climate policy necessarily makes energy more expensive, by a little for lenient emission reduction targets and smart policy design, and perhaps by a lot when targets are stringent and policies suboptimal. Energy is a necessary good; the burden of higher energy prices therefore falls disproportionally on the poor. However, the substitution away from fossil fuels reduces the return on capital and increases the demand for labour and wages. Climate policy may thus be a relative benefit to the working poor \citep{Rausch2011, Cullenward2016, Rausch2016, Melnikov2017, RosasFlores2017, TovarReanos2018, Goulder2019, Metcalf2019, Pizer2019, Saelim2019, Bohringer2021, Chepeliev2021, Faehn2021, Garaffa2021, Landis2021, Mayer2021, Vandyck2021, GarciaMuros2022, Wu2022}.

Estimates of the costs of climate policy vary widely between studies\textemdash predicting the future is hard\textemdash but all agree that a uniform carbon tax leading to the complete decarbonization of the economy by 2100 would be cheap, perhaps even too cheap to meter \citep{Clarke2014, Riahi2022IPCC}. The costs can be reduced by clever use of the revenues from carbon taxes and emission permit auctions \citep{Goulder1995}\textemdash if those policy instruments are indeed used. However, costs rapidly increase if policy is suboptimal\textemdash multiple emission permit markets, overlapping regulations such as a tax on top of tradable permits, unpredictably fluctuating subsidies, or inappropriate technical standards \citep{Boehringer2009}. Costs also increase rapidly if decarbonization needs to be completed well before 2100.

Climate policy affects energy access as well. Under pressure from donor countries and climate activists, development banks and, increasingly, investment banks have stopped the financing of fossil fuel projects in developing countries.\footnote{See \href{https://www.worldbank.org/en/topic/extractiveindustries/justtransition}{World Bank}, \href{https://www.adb.org/what-we-do/energy-policy}{Asian Development Bank}, and \href{https://www.hsbc.com/news-and-media/hsbc-news/were-phasing-out-coal-financing}{HSBC}.}
In many parts of the world, coal-fired power is the cheapest. Restricting investment and driving up the price of electricity reduces access, excluding more people and companies from using electricity and the appliances that use electricity.

\section{Discussion and conclusion}
\label{sc:conclude}
Measuring energy security is difficult, mostly because it consists of two conflicting parts, energy reliability and energy affordability, both of which are easy to measure \textit{ex post} but harder to predict \textit{ex ante}.

Providing energy security requires state intervention. Peak capacity is a public good, best purchased in a reverse auction. Redundancy in transport and transmission network monopolies is best achieved by direct regulation. Policies on energy access and poverty should focus on well-targeted income support and investment subsidies.

Because the supply of fossil fuels is spatially concentrated, political unrest in the areas of production can have worldwide effects. Outside interference may increase instability, as may the resource rents from oil and gas exploitation.

In the long run, climate policy and the replacement of fossil fuels with renewable energy should reduce the geopolitical risks to the energy supply. In the medium term, however, climate policy reduces energy affordability and, through asset stranding and bankruptcy, may negatively affect reliability too.

As illustrated by the many references above, there is a vast amount of research on energy security. Further research would be welcome in some areas. Indicators of the reliability of the electricity supply are readily available, but lacking for other energy sources and carriers. Internationally comparable indicators of energy access can be found, but not of energy poverty. Research on policy interventions to increase energy access and reduce energy poverty would proceed most fruitfully via field experiments, in which energy companies, regulators, and academics collaborate to test which policies work well and which not so well. Improved quantification of the probability of the outbreak of violent conflict would be a great boon. We do not understand enough about the impact of asset stranding and second-best climate policies.

Policy implications are implied in the above. The key policy recommendation, however, follows from the reality of the energy trilemma\textemdash the impossibility of energy that is clean, reliable, and cheap. If policymakers push too hard on one dimension of the energy trilemma, the other two will suffer. Energy policy, therefore, requires a careful and balanced consideration of all options.

\bibliography{master}

\begin{thebibliography}{256}
\providecommand{\natexlab}[1]{#1}
\providecommand{\url}[1]{\texttt{#1}}
\expandafter\ifx\csname urlstyle\endcsname\relax
  \providecommand{\doi}[1]{doi: #1}\else
  \providecommand{\doi}{doi: \begingroup \urlstyle{rm}\Url}\fi

\bibitem[Abeberese(2017)]{Abeberese2017}
A.~Abeberese.
\newblock Electricity cost and firm performance: Evidence from {I}ndia.
\newblock \emph{Review of Economics and Statistics}, 99\penalty0 (5):\penalty0
  839--852, 2017.
\newblock \doi{10.1162/REST_a_00641}.

\bibitem[Acharya and Sadath(2022)]{Acharya2022}
R.~Acharya and A.~Sadath.
\newblock Achievements and challenges of energy poverty alleviation policies:
  Evidence from the select states in {I}ndia.
\newblock \emph{Journal of Public Affairs}, 2022.
\newblock \doi{10.1002/pa.2839}.

\bibitem[Acheampong et~al.(2021{\natexlab{a}})Acheampong, Dzator, and
  Shahbaz]{Acheampong2021EE}
A.~Acheampong, J.~Dzator, and M.~Shahbaz.
\newblock Empowering the powerless: Does access to energy improve income
  inequality?
\newblock \emph{Energy Economics}, 99, 2021{\natexlab{a}}.
\newblock \doi{10.1016/j.eneco.2021.105288}.

\bibitem[Acheampong et~al.(2021{\natexlab{b}})Acheampong, Erdiaw-Kwasie, and
  Abunyewah]{Acheampong2021}
A.~Acheampong, M.~Erdiaw-Kwasie, and M.~Abunyewah.
\newblock Does energy accessibility improve human development? evidence from
  energy-poor regions.
\newblock \emph{Energy Economics}, 96, 2021{\natexlab{b}}.
\newblock \doi{10.1016/j.eneco.2021.105165}.

\bibitem[Adom and Nsabimana(2022)]{Adom2022}
P.~Adom and A.~Nsabimana.
\newblock Rural access to electricity and welfare outcomes in {R}wanda:
  Addressing issues of transitional heterogeneities and between and within
  gender disparities.
\newblock \emph{Resource and Energy Economics}, 70, 2022.
\newblock \doi{10.1016/j.reseneeco.2022.101333}.

\bibitem[Ahearne(2011)]{Ahearne2011}
J.~Ahearne.
\newblock Prospects for nuclear energy.
\newblock \emph{Energy Economics}, 33\penalty0 (4):\penalty0 572--580, 2011.
\newblock \doi{10.1016/j.eneco.2010.11.014}.

\bibitem[Aklin et~al.(2018)Aklin, Bayer, Harish, and Urpelainen]{Aklin2018}
M.~Aklin, P.~Bayer, S.~Harish, and J.~Urpelainen.
\newblock Economics of household technology adoption in developing countries:
  Evidence from solar technology adoption in rural {I}ndia.
\newblock \emph{Energy Economics}, 72:\penalty0 35--46, 2018.
\newblock \doi{10.1016/j.eneco.2018.02.011}.

\bibitem[Akpandjar and Kitchens(2017)]{Akpandjar2017}
G.~Akpandjar and C.~Kitchens.
\newblock From darkness to light: The effect of electrification in {G}hana,
  2000-2010.
\newblock \emph{Economic Development and Cultural Change}, 66\penalty0
  (1):\penalty0 31--54, 2017.
\newblock \doi{10.1086/693707}.

\bibitem[Alberini et~al.(2022)Alberini, Steinbuks, and Timilsina]{Alberini2022}
A.~Alberini, J.~Steinbuks, and G.~Timilsina.
\newblock How valuable is the reliability of residential electricity supply in
  low-income countries? evidence from {N}epal.
\newblock \emph{Energy Journal}, 43\penalty0 (4):\penalty0 1--26, 2022.
\newblock \doi{10.5547/01956574.43.4.aalb}.

\bibitem[Alby et~al.(2013)Alby, Dethier, and Straub]{Alby2013}
P.~Alby, J.~Dethier, and S.~Straub.
\newblock Firms operating under electricity constraints in developing
  countries.
\newblock \emph{World Bank Economic Review}, 27\penalty0 (1):\penalty0
  109--132, 2013.
\newblock \doi{10.1093/wber/lhs018}.

\bibitem[Alexeev and Conrad(2009)]{Alexeev2009}
M.~Alexeev and R.~Conrad.
\newblock The elusive curse of oil.
\newblock \emph{Review of Economics and Statistics}, 91\penalty0 (3):\penalty0
  586--598, 2009.
\newblock \doi{10.1162/rest.91.3.586}.

\bibitem[Allcott et~al.(2016)Allcott, Collard-Wexler, and
  O'Connell]{Allcott2016}
H.~Allcott, A.~Collard-Wexler, and S.~D. O'Connell.
\newblock How do electricity shortages affect industry? evidence from india.
\newblock \emph{American Economic Review}, 106\penalty0 (3):\penalty0 587--624,
  2016.
\newblock \doi{10.1257/aer.20140389}.

\bibitem[Amador et~al.(2013)Amador, González, and Ramos-Real]{Amador2013}
F.~J. Amador, R.~M. González, and F.~J. Ramos-Real.
\newblock Supplier choice and {WTP} for electricity attributes in an emerging
  market: The role of perceived past experience, environmental concern and
  energy saving behavior.
\newblock \emph{Energy Economics}, 40:\penalty0 953--966, 2013.
\newblock \doi{10.1016/j.eneco.2013.06.007}.

\bibitem[Ampofo and Mabefam(2021)]{Ampofo2021}
A.~Ampofo and M.~Mabefam.
\newblock Religiosity and energy poverty: Empirical evidence across countries.
\newblock \emph{Energy Economics}, 102, 2021.
\newblock \doi{10.1016/j.eneco.2021.105463}.

\bibitem[Andadari et~al.(2014)Andadari, Mulder, and Rietveld]{Andadari2014}
R.~Andadari, P.~Mulder, and P.~Rietveld.
\newblock Energy poverty reduction by fuel switching. impact evaluation of the
  {LPG} conversion program in {I}ndonesia.
\newblock \emph{Energy Policy}, 66:\penalty0 436--449, 2014.
\newblock \doi{10.1016/j.enpol.2013.11.021}.

\bibitem[Andersen and Dalgaard(2013)]{Andersen2013}
T.~Andersen and C.-J. Dalgaard.
\newblock Power outages and economic growth in africa.
\newblock \emph{Energy Economics}, 38:\penalty0 19--23, 2013.
\newblock \doi{10.1016/j.eneco.2013.02.016}.

\bibitem[Anderson(2019)]{Anderson2019}
D.~Anderson.
\newblock \emph{Environmental Economics and Natural Resource Management (5th
  edition)}.
\newblock Routledge, 2019.

\bibitem[Ang et~al.(2015)Ang, Choong, and Ng]{Ang2015}
B.~W. Ang, W.~L. Choong, and T.~S. Ng.
\newblock Energy security: Definitions, dimensions and indexes.
\newblock \emph{Renewable and Sustainable Energy Reviews}, 42:\penalty0
  1077--1093, 2015.
\newblock \doi{10.1016/j.rser.2014.10.064}.

\bibitem[Apergis et~al.(2022)Apergis, Polemis, and Soursou]{Apergis2022}
N.~Apergis, M.~Polemis, and S.-E. Soursou.
\newblock Energy poverty and education: Fresh evidence from a panel of
  developing countries.
\newblock \emph{Energy Economics}, 106, 2022.
\newblock \doi{10.1016/j.eneco.2021.105430}.

\bibitem[Aslaksen(2010)]{Aslaksen2010}
S.~Aslaksen.
\newblock Oil and democracy: More than a cross-country correlation?
\newblock \emph{Journal of Peace Research}, 47\penalty0 (4):\penalty0 421--431,
  2010.
\newblock \doi{10.1177/0022343310368348}.

\bibitem[Awaworyi~Churchill and Smyth(2020)]{AwaworyiChurchill2020ethnic}
S.~Awaworyi~Churchill and R.~Smyth.
\newblock Ethnic diversity, energy poverty and the mediating role of trust:
  Evidence from household panel data for {A}ustralia.
\newblock \emph{Energy Economics}, 86, 2020.
\newblock \doi{10.1016/j.eneco.2020.104663}.

\bibitem[Awaworyi~Churchill and Smyth(2021)]{AwaworyiChurchill2021}
S.~Awaworyi~Churchill and R.~Smyth.
\newblock Energy poverty and health: Panel data evidence from {A}ustralia.
\newblock \emph{Energy Economics}, 97, 2021.
\newblock \doi{10.1016/j.eneco.2021.105219}.

\bibitem[Awaworyi~Churchill and
  Smyth(2022{\natexlab{a}})]{AwaworyiChurchill2022crime}
S.~Awaworyi~Churchill and R.~Smyth.
\newblock Local area crime and energy poverty.
\newblock \emph{Energy Economics}, 114, 2022{\natexlab{a}}.
\newblock \doi{10.1016/j.eneco.2022.106274}.

\bibitem[Awaworyi~Churchill and
  Smyth(2022{\natexlab{b}})]{AwaworyiChurchill2022religion}
S.~Awaworyi~Churchill and R.~Smyth.
\newblock Protestantism and energy poverty.
\newblock \emph{Energy Economics}, 111, 2022{\natexlab{b}}.
\newblock \doi{10.1016/j.eneco.2022.106087}.

\bibitem[Awaworyi~Churchill et~al.(2020)Awaworyi~Churchill, Smyth, and
  Farrell]{AwaworyiChurchill2020welfare}
S.~Awaworyi~Churchill, R.~Smyth, and L.~Farrell.
\newblock Fuel poverty and subjective wellbeing.
\newblock \emph{Energy Economics}, 86, 2020.
\newblock \doi{10.1016/j.eneco.2019.104650}.

\bibitem[Aweke and Navrud(2022)]{Aweke2022}
A.~T. Aweke and S.~Navrud.
\newblock Valuing energy poverty costs: Household welfare loss from electricity
  blackouts in developing countries.
\newblock \emph{Energy Economics}, 109, 2022.
\newblock \doi{10.1016/j.eneco.2022.105943}.

\bibitem[Ayana and Degaga(2022)]{Ayana2022}
O.~Ayana and J.~Degaga.
\newblock Effects of rural electrification on household welfare: a
  meta-regression analysis.
\newblock \emph{International Review of Economics}, 69\penalty0 (2):\penalty0
  209--261, 2022.
\newblock \doi{10.1007/s12232-022-00391-7}.

\bibitem[Badeeb et~al.(2017)Badeeb, Lean, and Clark]{Badeeb2017}
R.~Badeeb, H.~Lean, and J.~Clark.
\newblock The evolution of the natural resource curse thesis: A critical
  literature survey.
\newblock \emph{Resources Policy}, 51:\penalty0 123--134, 2017.
\newblock \doi{10.1016/j.resourpol.2016.10.015}.

\bibitem[Bagnoli and Bertom\'{e}u-S\'{a}nchez(2022)]{Bagnoli2022}
L.~Bagnoli and S.~Bertom\'{e}u-S\'{a}nchez.
\newblock How effective has the electricity social rate been in reducing energy
  poverty in spain?
\newblock \emph{Energy Economics}, 106, 2022.
\newblock \doi{10.1016/j.eneco.2021.105792}.

\bibitem[Bajo-Buenestado(2021)]{Bajo-Buenestado2021}
R.~Bajo-Buenestado.
\newblock The effect of blackouts on household electrification status: Evidence
  from {K}enya.
\newblock \emph{Energy Economics}, 94, 2021.
\newblock \doi{10.1016/j.eneco.2020.105067}.

\bibitem[Barnes et~al.(2011)Barnes, Khandker, and Samad]{Barnes2011}
D.~Barnes, S.~Khandker, and H.~Samad.
\newblock Energy poverty in rural {B}angladesh.
\newblock \emph{Energy Policy}, 39\penalty0 (2):\penalty0 894--904, 2011.
\newblock \doi{10.1016/j.enpol.2010.11.014}.

\bibitem[Barron and Torero(2017)]{Barron2017}
M.~Barron and M.~Torero.
\newblock Household electrification and indoor air pollution.
\newblock \emph{Journal of Environmental Economics and Management},
  86:\penalty0 81--92, 2017.
\newblock \doi{10.1016/j.jeem.2017.07.007}.

\bibitem[Basedau and Lay(2009)]{Basedau2009}
M.~Basedau and J.~Lay.
\newblock Resource curse or rentier peace? the ambiguous effects of oil wealth
  and oil dependence on violent conflict.
\newblock \emph{Journal of Peace Research}, 46\penalty0 (6):\penalty0 757--776,
  2009.
\newblock \doi{10.1177/0022343309340500}.

\bibitem[Baumol and Oates(1988)]{Baumol1988}
W.~Baumol and W.~Oates.
\newblock \emph{The theory of environmental policy (2nd edition)}.
\newblock Cambridge University Press, 1988.

\bibitem[Beenstock(1991)]{Beenstock1991}
M.~Beenstock.
\newblock Generators and the cost of electricity outages.
\newblock \emph{Energy Economics}, 13\penalty0 (4):\penalty0 283--289, 1991.
\newblock \doi{10.1016/0140-9883(91)90008-N}.

\bibitem[Beenstock et~al.(1997)Beenstock, Goldin, and Haitovsky]{Beenstock1997}
M.~Beenstock, E.~Goldin, and Y.~Haitovsky.
\newblock The cost of power outages in the business and public sectors in
  {I}srael: Revealed preference vs. subjective valuation.
\newblock \emph{Energy Journal}, 18\penalty0 (2):\penalty0 39--61, 1997.
\newblock \doi{10.5547/ISSN0195-6574-EJ-Vol18-No2-3}.

\bibitem[Bell and Davis(2001)]{Bell2001}
M.~Bell and D.~Davis.
\newblock Reassessment of the lethal {L}ondon fog of 1952: Novel indicators of
  acute and chronic consequences of acute exposure to air pollution.
\newblock \emph{Environmental Health Perspectives}, 109\penalty0 (SUPPL.
  3):\penalty0 389--394, 2001.
\newblock \doi{10.1289/ehp.01109s3389}.

\bibitem[Berck and Helfand(2011)]{Berck2011}
P.~Berck and G.~Helfand.
\newblock \emph{The Economics of the Environment}.
\newblock Addison Wesley, 2011.

\bibitem[Best et~al.(2021)Best, Hammerle, Mukhopadhaya, and Silber]{Best2021}
R.~Best, M.~Hammerle, P.~Mukhopadhaya, and J.~Silber.
\newblock Targeting household energy assistance.
\newblock \emph{Energy Economics}, 99, 2021.
\newblock \doi{10.1016/j.eneco.2021.105311}.

\bibitem[Betz et~al.(2015)Betz, Partridge, Farren, and Lobao]{Betz2015}
M.~Betz, M.~Partridge, M.~Farren, and L.~Lobao.
\newblock Coal mining, economic development, and the natural resources curse.
\newblock \emph{Energy Economics}, 50:\penalty0 105--116, 2015.
\newblock \doi{10.1016/j.eneco.2015.04.005}.

\bibitem[Bhattacharyya(2011)]{Bhattacharyya2011}
S.~C. Bhattacharyya.
\newblock \emph{Energy economics: Concepts, issues, markets and governance}.
\newblock 2011.
\newblock \doi{10.1007/978-0-85729-268-1}.

\bibitem[Bo et~al.(2022)Bo, Chen, and Liu]{Bo2022}
S.~Bo, T.~Chen, and C.~Liu.
\newblock Trade shocks, industrial growth, and electrification in early
  20th-century {C}hina.
\newblock \emph{Journal of Comparative Economics}, 50\penalty0 (3):\penalty0
  732--749, 2022.
\newblock \doi{10.1016/j.jce.2022.02.001}.

\bibitem[Boehringer et~al.(2009)Boehringer, Rutherford, and
  Tol]{Boehringer2009}
C.~Boehringer, T.~F. Rutherford, and R.~S.~J. Tol.
\newblock The {EU} 20/20/2020 targets: An overview of the {EMF22} assessment.
\newblock \emph{Energy Economics}, 31\penalty0 (SUPPL. 2):\penalty0 S268--S273,
  2009.
\newblock \doi{10.1016/j.eneco.2009.10.010}.

\bibitem[Bohi and Toman(1993)]{Bohi1993}
D.~Bohi and M.~Toman.
\newblock Energy security: externalities and policies.
\newblock \emph{Energy Policy}, 21\penalty0 (11):\penalty0 1093--1109, 1993.
\newblock \doi{10.1016/0301-4215(93)90260-M}.

\bibitem[Bohi and Toman(1996)]{Bohi1996}
D.~Bohi and M.~Toman.
\newblock \emph{The Economics of Energy Security}.
\newblock Kluwer, Norwell, 1996.
\newblock cited By 143.

\bibitem[B\"{o}hringer and Bortolamedi(2015)]{Bohringer2015}
C.~B\"{o}hringer and M.~Bortolamedi.
\newblock Sense and no(n)-sense of energy security indicators.
\newblock \emph{Ecological Economics}, 119:\penalty0 359--371, 2015.
\newblock \doi{10.1016/j.ecolecon.2015.09.020}.

\bibitem[B\"{o}hringer and Jochem(2007)]{Bohringer2007}
C.~B\"{o}hringer and P.~E.~P. Jochem.
\newblock Measuring the immeasurable\textemdash a survey of sustainability
  indices.
\newblock \emph{Ecological Economics}, 63\penalty0 (1):\penalty0 1--8, 2007.
\newblock \doi{10.1016/j.ecolecon.2007.03.008}.

\bibitem[B\"{o}hringer et~al.(2021)B\"{o}hringer, Rutherford, and
  Schneider]{Bohringer2021}
C.~B\"{o}hringer, T.~F. Rutherford, and J.~Schneider.
\newblock The incidence of {CO}\textsubscript{2} emissions pricing under
  alternative international market responses: A computable general equilibrium
  analysis for {G}ermany.
\newblock \emph{Energy Economics}, 101:\penalty0 105404, 2021.
\newblock URL
  \url{https://www.sciencedirect.com/science/article/pii/S0140988321003017}.

\bibitem[Boschini et~al.(2007)Boschini, Pettersson, and Roine]{Boschini2007}
A.~Boschini, J.~Pettersson, and J.~Roine.
\newblock Resource curse or not: A question of appropriability.
\newblock \emph{Scandinavian Journal of Economics}, 109\penalty0 (3):\penalty0
  593--617, 2007.
\newblock \doi{10.1111/j.1467-9442.2007.00509.x}.

\bibitem[Boschini et~al.(2013)Boschini, Pettersson, and Roine]{Boschini2013}
A.~Boschini, J.~Pettersson, and J.~Roine.
\newblock The resource curse and its potential reversal.
\newblock \emph{World Development}, 43:\penalty0 19--41, 2013.
\newblock \doi{10.1016/j.worlddev.2012.10.007}.

\bibitem[Bouzarovski et~al.(2012)Bouzarovski, Petrova, and
  Sarlamanov]{Bouzarovski2012}
S.~Bouzarovski, S.~Petrova, and R.~Sarlamanov.
\newblock Energy poverty policies in the {EU}: A critical perspective.
\newblock \emph{Energy Policy}, 49:\penalty0 76--82, 2012.
\newblock \doi{10.1016/j.enpol.2012.01.033}.

\bibitem[Brollo et~al.(2013)Brollo, Nannicini, Perotti, and
  Tabellini]{Brollo2013}
F.~Brollo, T.~Nannicini, R.~Perotti, and G.~Tabellini.
\newblock The political resource curse.
\newblock \emph{American Economic Review}, 103\penalty0 (5):\penalty0
  1759--1796, 2013.
\newblock \doi{10.1257/aer.103.5.1759}.

\bibitem[Brunnschweiler(2008)]{Brunnschweiler2008WD}
C.~Brunnschweiler.
\newblock Cursing the blessings? natural resource abundance, institutions, and
  economic growth.
\newblock \emph{World Development}, 36\penalty0 (3):\penalty0 399--419, 2008.
\newblock \doi{10.1016/j.worlddev.2007.03.004}.

\bibitem[Brunnschweiler and Bulte(2008)]{Brunnschweiler2008JEEM}
C.~Brunnschweiler and E.~Bulte.
\newblock The resource curse revisited and revised: A tale of paradoxes and red
  herrings.
\newblock \emph{Journal of Environmental Economics and Management}, 55\penalty0
  (3):\penalty0 248--264, 2008.
\newblock \doi{10.1016/j.jeem.2007.08.004}.

\bibitem[Brunnschweiler and Bulte(2009)]{Brunnschweiler2009}
C.~Brunnschweiler and E.~Bulte.
\newblock Natural resources and violent conflict: Resource abundance,
  dependence, and the onset of civil wars.
\newblock \emph{Oxford Economic Papers}, 61\penalty0 (4):\penalty0 651--674,
  2009.
\newblock \doi{10.1093/oep/gpp024}.

\bibitem[Bukari et~al.(2021)Bukari, Broermann, and Okai]{Bukari2021}
C.~Bukari, S.~Broermann, and D.~Okai.
\newblock Energy poverty and health expenditure: Evidence from {G}hana.
\newblock \emph{Energy Economics}, 103, 2021.
\newblock \doi{10.1016/j.eneco.2021.105565}.

\bibitem[Bulte et~al.(2005)Bulte, Damania, and Deacon]{Bulte2005}
E.~Bulte, R.~Damania, and R.~Deacon.
\newblock Resource intensity, institutions, and development.
\newblock \emph{World Development}, 33\penalty0 (7):\penalty0 1029--1044, 2005.
\newblock \doi{10.1016/j.worlddev.2005.04.004}.

\bibitem[Burgess et~al.(2020)Burgess, Ritchie, Shapland, and {Pielke
  Jr}]{Burgess2020}
M.~G. Burgess, J.~Ritchie, J.~Shapland, and R.~A. {Pielke Jr}.
\newblock {IPCC} baseline scenarios have over-projected co\textsubscript{2}
  emissions and economic growth.
\newblock \emph{Environmental Research Letters}, 16\penalty0 (1):\penalty0
  024027, 2020.
\newblock \doi{10.1088/1748-9326/abcdd2}.

\bibitem[Burke et~al.(2018)Burke, Stern, and Bruns]{Burke2018}
P.~Burke, D.~Stern, and S.~Bruns.
\newblock The impact of electricity on economic development: A macroeconomic
  perspective.
\newblock \emph{International Review of Environmental and Resource Economics},
  12\penalty0 (1):\penalty0 85--127, 2018.
\newblock \doi{10.1561/101.00000101}.

\bibitem[Carlsson and Martinsson(2007)]{Carlsson2007}
F.~Carlsson and P.~Martinsson.
\newblock Willingness to pay among swedish households to avoid power outages: A
  random parameter tobit model approach.
\newblock \emph{Energy Journal}, 28\penalty0 (1):\penalty0 75--89, 2007.
\newblock \doi{10.5547/ISSN0195-6574-EJ-Vol28-No1-4}.

\bibitem[Carlsson and Martinsson(2008)]{Carlsson2008}
F.~Carlsson and P.~Martinsson.
\newblock Does it matter when a power outage occurs?\textemdash a choice
  experiment study on the willingness to pay to avoid power outages.
\newblock \emph{Energy Economics}, 30\penalty0 (3):\penalty0 1232--1245, 2008.
\newblock \doi{10.1016/j.eneco.2007.04.001}.

\bibitem[Carlsson et~al.(2011)Carlsson, Martinsson, and Akay]{Carlsson2011}
F.~Carlsson, P.~Martinsson, and A.~Akay.
\newblock The effect of power outages and cheap talk on willingness to pay to
  reduce outages.
\newblock \emph{Energy Economics}, 33\penalty0 (5):\penalty0 790--798, 2011.
\newblock \doi{10.1016/j.eneco.2011.01.004}.

\bibitem[Carlsson et~al.(2021)Carlsson, Kataria, Lampi, and
  Martinsson]{Carlsson2021}
F.~Carlsson, M.~Kataria, E.~Lampi, and P.~Martinsson.
\newblock Past and present outage costs\textemdash a follow-up study of
  households’ willingness to pay to avoid power outages.
\newblock \emph{Resource and Energy Economics}, 64, 2021.
\newblock \doi{10.1016/j.reseneeco.2021.101216}.

\bibitem[Carranza and Meeks(2021)]{Carranza2021}
E.~Carranza and R.~Meeks.
\newblock Energy efficiency and electricity reliability.
\newblock \emph{Review of Economics and Statistics}, 103\penalty0 (3):\penalty0
  461--475, 2021.
\newblock \doi{10.1162/rest_a_00912}.

\bibitem[Cavalcanti et~al.(2011)Cavalcanti, Mohaddes, and
  Raissi]{Cavalcanti2011}
T.~Cavalcanti, K.~Mohaddes, and M.~Raissi.
\newblock Growth, development and natural resources: New evidence using a
  heterogeneous panel analysis.
\newblock \emph{Quarterly Review of Economics and Finance}, 51\penalty0
  (4):\penalty0 305--318, 2011.
\newblock \doi{10.1016/j.qref.2011.07.007}.

\bibitem[Chakravarty and Tavoni(2013)]{Chakravarty2013}
S.~Chakravarty and M.~Tavoni.
\newblock Energy poverty alleviation and climate change mitigation: Is there a
  trade off?
\newblock \emph{Energy Economics}, 40:\penalty0 S67--S73, 2013.
\newblock \doi{10.1016/j.eneco.2013.09.022}.

\bibitem[Chakravorty et~al.(2014)Chakravorty, Pelli, and
  Ural~Marchand]{Chakravorty2014}
U.~Chakravorty, M.~Pelli, and B.~Ural~Marchand.
\newblock Does the quality of electricity matter? evidence from rural india.
\newblock \emph{Journal of Economic Behavior and Organization}, 107\penalty0
  (PA):\penalty0 228--247, 2014.
\newblock \doi{10.1016/j.jebo.2014.04.011}.

\bibitem[Chaurey and Le(2022)]{Chaurey2022}
R.~Chaurey and D.~Le.
\newblock Infrastructure maintenance and rural economic activity: Evidence
  {f}rom india.
\newblock \emph{Journal of Public Economics}, 214, 2022.
\newblock \doi{10.1016/j.jpubeco.2022.104725}.

\bibitem[Chen et~al.(2022)Chen, Yan, Gong, Geng, and Yuan]{Chen2022}
H.~Chen, H.~Yan, K.~Gong, H.~Geng, and X.-C. Yuan.
\newblock Assessing the business interruption costs from power outages in
  {C}hina.
\newblock \emph{Energy Economics}, 105, 2022.
\newblock \doi{10.1016/j.eneco.2021.105757}.

\bibitem[Chepeliev et~al.(2021)Chepeliev, Osorio-Rodarte, and {van der
  Mensbrugghe}]{Chepeliev2021}
M.~Chepeliev, I.~Osorio-Rodarte, and D.~{van der Mensbrugghe}.
\newblock Distributional impacts of carbon pricing policies under the paris
  agreement: Inter and intra-regional perspectives.
\newblock \emph{Energy Economics}, 102:\penalty0 105530, 2021.
\newblock URL
  \url{https://www.sciencedirect.com/science/article/pii/S0140988321004084}.

\bibitem[Cherp and Jewell(2011)]{Cherp2011}
A.~Cherp and J.~Jewell.
\newblock Measuring energy security: from universal indicators to
  contextualized frameworks.
\newblock \emph{The Routledge Handbook of Energy Security}, pages 330--355,
  2011.

\bibitem[Chhay and Yamazaki(2021)]{Chhay2021}
P.~Chhay and K.~Yamazaki.
\newblock Rural electrification and changes in employment structure in
  {C}ambodia.
\newblock \emph{World Development}, 137, 2021.
\newblock \doi{10.1016/j.worlddev.2020.105212}.

\bibitem[Clarke et~al.(2014)Clarke, Jiang, Akimoto, Babiker, Blanford,
  Fisher-Vanden, Hourcade, Krey, Kriegler, Loeschel, McCollum, Paltsev, Rose,
  Shukla, Tavoni, van Vuuren, and Van Der~Zwaan]{Clarke2014}
L.~Clarke, K.~Jiang, K.~Akimoto, M.~H. Babiker, G.~J. Blanford, K.~A.
  Fisher-Vanden, J.~C. Hourcade, V.~Krey, E.~Kriegler, A.~Loeschel, D.~W.
  McCollum, S.~Paltsev, S.~Rose, P.~R. Shukla, M.~Tavoni, D.~van Vuuren, and
  B.~Van Der~Zwaan.
\newblock Assessing transformation pathways.
\newblock In O.~Edenhofer, R.~Pichs-Madruga, and Y.~Sokona, editors,
  \emph{Climate Change 2014: Mitigation of Climate Change\textemdash
  Contribution of Working Group III to the Fifth Assessment Report of the
  Intergovernmental Panel on Climate Change}. Cambridge University Press,
  Cambridge, 2014.

\bibitem[Cole et~al.(2018)Cole, Elliott, Occhiali, and Strobl]{Cole2018}
M.~A. Cole, R.~J.~R. Elliott, G.~Occhiali, and E.~Strobl.
\newblock Power outages and firm performance in {Sub-Saharan Africa}.
\newblock \emph{Journal of Development Economics}, 134:\penalty0 150--159,
  2018.
\newblock \doi{10.1016/j.jdeveco.2018.05.003}.

\bibitem[Collier and Hoeffler(2005)]{Collier2005}
P.~Collier and A.~Hoeffler.
\newblock Resource rents, governance, and conflict.
\newblock \emph{Journal of Conflict Resolution}, 49\penalty0 (4):\penalty0
  625--633, 2005.
\newblock \doi{10.1177/0022002705277551}.

\bibitem[Cravioto et~al.(2020)Cravioto, Ohgaki, Che, Tan, Kobayashi, Toe, Long,
  Oudaya, Rahim, and Farzeneh]{Cravioto2020}
J.~Cravioto, H.~Ohgaki, H.~Che, C.~Tan, S.~Kobayashi, H.~Toe, B.~Long,
  E.~Oudaya, N.~Rahim, and H.~Farzeneh.
\newblock The effects of rural electrification on quality of life: A {Southeast
  Asian} perspective.
\newblock \emph{Energies}, 13\penalty0 (10), 2020.
\newblock \doi{10.3390/en13102410}.

\bibitem[Crentsil et~al.(2019)Crentsil, Asuman, and Fenny]{Crentsil2019}
A.~Crentsil, D.~Asuman, and A.~Fenny.
\newblock Assessing the determinants and drivers of multidimensional energy
  poverty in {G}hana.
\newblock \emph{Energy Policy}, 133, 2019.
\newblock \doi{10.1016/j.enpol.2019.110884}.

\bibitem[Creti and Fabra(2007)]{Creti2007}
A.~Creti and N.~Fabra.
\newblock Supply security and short-run capacity markets for electricity.
\newblock \emph{Energy Economics}, 29\penalty0 (2):\penalty0 259--276, 2007.
\newblock \doi{10.1016/j.eneco.2006.04.007}.

\bibitem[Cullenward et~al.(2016)Cullenward, Wilkerson, Wara, and
  Weyant]{Cullenward2016}
D.~Cullenward, J.~Wilkerson, M.~Wara, and J.~Weyant.
\newblock Dynamically estimating the distributional impacts of {U.S.} climate
  policy with {NEMS}: A case study of the {Climate Protection Act} of 2013.
\newblock \emph{Energy Economics}, 55:\penalty0 303--318, 2016.
\newblock \doi{10.1016/j.eneco.2016.02.021}.

\bibitem[Da~Silveira~Bezerra et~al.(2017)Da~Silveira~Bezerra, Callegari, Ribas,
  Lucena, Portugal-Pereira, Koberle, Szklo, and
  Schaeffer]{DaSilveiraBezerra2017}
P.~Da~Silveira~Bezerra, C.~Callegari, A.~Ribas, A.~Lucena, J.~Portugal-Pereira,
  A.~Koberle, A.~Szklo, and R.~Schaeffer.
\newblock The power of light: Socio-economic and environmental implications of
  a rural electrification program in {B}razil.
\newblock \emph{Environmental Research Letters}, 12\penalty0 (9), 2017.
\newblock \doi{10.1088/1748-9326/aa7bdd}.

\bibitem[Dang and La(2019)]{Dang2019}
D.~Dang and H.~La.
\newblock Does electricity reliability matter? evidence from rural {Viet Nam}.
\newblock \emph{Energy Policy}, 131:\penalty0 399--409, 2019.
\newblock \doi{10.1016/j.enpol.2019.04.036}.

\bibitem[Dasso and Fernandez(2015)]{Dasso2015}
R.~Dasso and F.~Fernandez.
\newblock The effects of electrification on employment in rural {P}eru.
\newblock \emph{IZA Journal of Labor and Development}, 4\penalty0 (1), 2015.
\newblock \doi{10.1186/s40175-015-0028-4}.

\bibitem[Davis et~al.(2010)Davis, Caldeira, and Matthews]{Davis2010}
S.~J. Davis, K.~Caldeira, and H.~D. Matthews.
\newblock Future {CO}\textsubscript{2} emissions and climate change from
  existing energy infrastructure.
\newblock \emph{Science}, 329\penalty0 (5997):\penalty0 1330--1333, 2010.
\newblock \doi{10.1126/science.1188566}.

\bibitem[de~Nooij et~al.(2007)de~Nooij, Koopmans, and Bijvoet]{deNooij2007}
M.~de~Nooij, C.~Koopmans, and C.~Bijvoet.
\newblock The value of supply security. the costs of power interruptions:
  Economic input for damage reduction and investment in networks.
\newblock \emph{Energy Economics}, 29\penalty0 (2):\penalty0 277--295, 2007.
\newblock \doi{10.1016/j.eneco.2006.05.022}.

\bibitem[de~Nooij et~al.(2009)de~Nooij, Lieshout, and Koopmans]{deNooij2009}
M.~de~Nooij, R.~Lieshout, and C.~Koopmans.
\newblock Optimal blackouts: Empirical results on reducing the social cost of
  electricity outages through efficient regional rationing.
\newblock \emph{Energy Economics}, 31\penalty0 (3):\penalty0 342--347, 2009.
\newblock \doi{10.1016/j.eneco.2008.11.004}.

\bibitem[Dendup(2022)]{Dendup2022}
N.~Dendup.
\newblock Returns to grid electricity on firewood and kerosene: Mechanism.
\newblock \emph{Journal of Environmental Economics and Management}, 111, 2022.
\newblock \doi{10.1016/j.jeem.2021.102606}.

\bibitem[Deutschmann et~al.(2021)Deutschmann, Postepska, and
  Sarr]{Deutschmann2021}
J.~Deutschmann, A.~Postepska, and L.~Sarr.
\newblock Measuring willingness to pay for reliable electricity: Evidence from
  {S}enegal.
\newblock \emph{World Development}, 138, 2021.
\newblock \doi{10.1016/j.worlddev.2020.105209}.

\bibitem[Diallo and Moussa(2020)]{Diallo2020}
A.~Diallo and R.~Moussa.
\newblock The effects of solar home system on welfare in off-grid areas:
  Evidence from {C}\^{o}te {d'I}voire.
\newblock \emph{Energy}, 194, 2020.
\newblock \doi{10.1016/j.energy.2019.116835}.

\bibitem[Ding et~al.(2018)Ding, Qin, and Shi]{Ding2018}
H.~Ding, C.~Qin, and K.~Shi.
\newblock Development through electrification: Evidence from rural {C}hina.
\newblock \emph{China Economic Review}, 50:\penalty0 313--328, 2018.
\newblock \doi{10.1016/j.chieco.2018.04.007}.

\bibitem[Dinkelman(2011)]{Dinkelman2011}
T.~Dinkelman.
\newblock The effects of rural electrification on employment: New evidence from
  south africa.
\newblock \emph{American Economic Review}, 101\penalty0 (7):\penalty0
  3078--3108, 2011.
\newblock \doi{10.1257/aer.101.7.3078}.

\bibitem[Dogan et~al.(2022)Dogan, Madaleno, Inglesi-Lotz, and
  Taskin]{Dogan2022}
E.~Dogan, M.~Madaleno, R.~Inglesi-Lotz, and D.~Taskin.
\newblock Race and energy poverty: Evidence from african-american households.
\newblock \emph{Energy Economics}, 108, 2022.
\newblock \doi{10.1016/j.eneco.2022.105908}.

\bibitem[Dunning(2005)]{Dunning2005}
T.~Dunning.
\newblock Resource dependence, economic performance, and political stability.
\newblock \emph{Journal of Conflict Resolution}, 49\penalty0 (4):\penalty0
  451--482, 2005.
\newblock \doi{10.1177/0022002705277521}.

\bibitem[Elliott et~al.(2021)Elliott, Nguyen-Tien, and Strobl]{Elliott2021}
R.~J.~R. Elliott, V.~Nguyen-Tien, and E.~A. Strobl.
\newblock Power outages and firm performance: A hydro-{IV} approach for a
  single electricity grid.
\newblock \emph{Energy Economics}, 103, 2021.
\newblock \doi{10.1016/j.eneco.2021.105571}.

\bibitem[Emmanuel and Japhet(2020)]{Emmanuel2020}
N.~Emmanuel and K.~Japhet.
\newblock Impact of rural electrification on {U}gandan women’s empowerment:
  Evidence from micro-data.
\newblock \emph{Journal of Energy and Development}, 46\penalty0 (1):\penalty0
  63--80, 2020.
\newblock \doi{10.2307/27107169}.

\bibitem[Endres and Radke(2012)]{Endres2012}
A.~Endres and V.~Radke.
\newblock \emph{Economics for Environmental Studies}.
\newblock Heidelberg, 2012.

\bibitem[Feeny et~al.(2021)Feeny, Trinh, and Zhu]{Feeny2021}
S.~Feeny, T.-A. Trinh, and A.~Zhu.
\newblock Temperature shocks and energy poverty: Findings from {V}ietnam.
\newblock \emph{Energy Economics}, 99, 2021.
\newblock \doi{10.1016/j.eneco.2021.105310}.

\bibitem[Field and Field(2009)]{Field2009}
B.~Field and M.~Field.
\newblock \emph{Environmental Economics\textemdash An Introduction (5th
  edition)}.
\newblock New York, 2009.

\bibitem[Fried and Lagakos(2021)]{Fried2021}
S.~Fried and D.~Lagakos.
\newblock Rural electrification, migration and structural transformation:
  Evidence from {E}thiopia.
\newblock \emph{Regional Science and Urban Economics}, 91, 2021.
\newblock \doi{10.1016/j.regsciurbeco.2020.103625}.

\bibitem[Frondel and Schmidt(2014)]{Frondel2014}
M.~Frondel and C.~M. Schmidt.
\newblock A measure of a nation's physical energy supply risk.
\newblock \emph{Quarterly Review of Economics and Finance}, 54\penalty0
  (2):\penalty0 208--215, 2014.
\newblock \doi{10.1016/j.qref.2013.10.003}.

\bibitem[Fujii and Shonchoy(2020)]{Fujii2020}
T.~Fujii and A.~Shonchoy.
\newblock Fertility and rural electrification in {B}angladesh.
\newblock \emph{Journal of Development Economics}, 143, 2020.
\newblock \doi{10.1016/j.jdeveco.2019.102430}.

\bibitem[Fujii et~al.(2018)Fujii, Shonchoy, and Xu]{Fujii2018}
T.~Fujii, A.~Shonchoy, and S.~Xu.
\newblock Impact of electrification on children's nutritional status in rural
  {B}angladesh.
\newblock \emph{World Development}, 102:\penalty0 315--330, 2018.
\newblock \doi{10.1016/j.worlddev.2017.07.016}.

\bibitem[Gafa and Egbendewe(2021)]{Gafa2021}
D.~Gafa and A.~Egbendewe.
\newblock Energy poverty in rural {West Africa} and its determinants: Evidence
  from {S}enegal and {T}ogo.
\newblock \emph{Energy Policy}, 156, 2021.
\newblock \doi{10.1016/j.enpol.2021.112476}.

\bibitem[Gaggl et~al.(2021)Gaggl, Gray, Marinescu, and Morin]{Gaggl2021}
P.~Gaggl, R.~Gray, I.~Marinescu, and M.~Morin.
\newblock Does electricity drive structural transformation? evidence from the
  {United States}.
\newblock \emph{Labour Economics}, 68, 2021.
\newblock \doi{10.1016/j.labeco.2020.101944}.

\bibitem[Garaffa et~al.(2021)Garaffa, Cunha, Cruz, Bezerra, Lucena, and
  Gurgel]{Garaffa2021}
R.~Garaffa, B.~S.~L. Cunha, T.~Cruz, P.~Bezerra, A.~F.~P. Lucena, and A.~C.
  Gurgel.
\newblock Distributional effects of carbon pricing in {B}razil under the
  {P}aris {A}greement.
\newblock \emph{Energy Economics}, 101:\penalty0 105396, 2021.
\newblock URL
  \url{https://www.sciencedirect.com/science/article/pii/S0140988321002954}.

\bibitem[Garc\'{i}a~Alvarez and Tol(2021)]{GarciaAlvarez2021}
G.~Garc\'{i}a~Alvarez and R.~S.~J. Tol.
\newblock The impact of the {Bono Social de Electricidad} on energy poverty in
  {S}pain.
\newblock \emph{Energy Economics}, 103, 2021.
\newblock \doi{10.1016/j.eneco.2021.105554}.

\bibitem[Garc\'{i}a-Muros et~al.(2022)Garc\'{i}a-Muros, Morris, and
  Paltsev]{GarciaMuros2022}
X.~Garc\'{i}a-Muros, J.~Morris, and S.~Paltsev.
\newblock Toward a just energy transition: A distributional analysis of
  low-carbon policies in the usa.
\newblock \emph{Energy Economics}, 105:\penalty0 105769, 2022.
\newblock URL
  \url{https://www.sciencedirect.com/science/article/pii/S0140988321006113}.

\bibitem[Garc\'{i}a-Verdugo and Munoz(2012)]{Garcia2012}
J.~Garc\'{i}a-Verdugo and B.~Munoz.
\newblock Energy dependence, vulnerability and the geopolitical context: A
  quantitative approach to energy security.
\newblock pages 37--53. 2012.
\newblock \doi{10.4324/9780203153291}.

\bibitem[Gnansounou(2008)]{Gnansounou2008}
E.~Gnansounou.
\newblock Assessing the energy vulnerability: Case of industrialised countries.
\newblock \emph{Energy Policy}, 36\penalty0 (10):\penalty0 3734--3744, 2008.
\newblock \doi{10.1016/j.enpol.2008.07.004}.

\bibitem[Goodstein(2005)]{Goodstein2005}
E.~Goodstein.
\newblock \emph{Economics and the Environment (4th edition)}.
\newblock Hoboken, 2005.

\bibitem[Goulder(1995)]{Goulder1995}
L.~Goulder.
\newblock Environmental taxation and the double dividend: A reader's guide.
\newblock \emph{International Tax and Public Finance}, 2\penalty0 (2):\penalty0
  157--183, 1995.
\newblock \doi{10.1007/BF00877495}.

\bibitem[Goulder et~al.(2019)Goulder, Hafstead, Kim, and Long]{Goulder2019}
L.~H. Goulder, M.~A.~C. Hafstead, G.~Kim, and X.~Long.
\newblock Impacts of a carbon tax across us household income groups: What are
  the equity-efficiency trade-offs?
\newblock \emph{Journal of Public Economics}, 175:\penalty0 44--64, 2019.
\newblock \doi{10.1016/j.jpubeco.2019.04.002}.

\bibitem[Gracceva and Zeniewski(2014)]{Gracceva2014}
F.~Gracceva and P.~Zeniewski.
\newblock A systemic approach to assessing energy security in a low-carbon eu
  energy system.
\newblock \emph{Applied Energy}, 123:\penalty0 335?348, 2014.
\newblock \doi{10.1016/j.apenergy.2013.12.018}.

\bibitem[Grimm et~al.(2015)Grimm, Sparrow, and Tasciotti]{Grimm2015}
M.~Grimm, R.~Sparrow, and L.~Tasciotti.
\newblock Does electrification spur the fertility transition? evidence from
  {I}ndonesia.
\newblock \emph{Demography}, 52\penalty0 (5):\penalty0 1773--1796, 2015.
\newblock \doi{10.1007/s13524-015-0420-3}.

\bibitem[Grimm et~al.(2017)Grimm, Munyehirwe, Peters, and Sievert]{Grimm2017}
M.~Grimm, A.~Munyehirwe, J.~Peters, and M.~Sievert.
\newblock A first step up the energy ladder? low cost solar kits and
  household's welfare in rural {R}wanda.
\newblock \emph{World Bank Economic Review}, 31\penalty0 (3):\penalty0
  631--649, 2017.
\newblock \doi{10.1093/wber/lhw052}.

\bibitem[Grogan(2016)]{Grogan2016}
L.~Grogan.
\newblock Household electrification, fertility, and employment: Evidence from
  hydroelectric dam construction in {C}olombia.
\newblock \emph{Journal of Human Capital}, 10\penalty0 (1):\penalty0 109--158,
  2016.
\newblock \doi{10.1086/684580}.

\bibitem[Grogan(2018)]{Grogan2018}
L.~Grogan.
\newblock Time use impacts of rural electrification: Longitudinal evidence from
  {G}uatemala.
\newblock \emph{Journal of Development Economics}, 135:\penalty0 304--317,
  2018.
\newblock \doi{10.1016/j.jdeveco.2018.03.005}.

\bibitem[Grogan and Sadanand(2013)]{Grogan2013}
L.~Grogan and A.~Sadanand.
\newblock Rural electrification and employment in poor countries: Evidence from
  {N}icaragua.
\newblock \emph{World Development}, 43:\penalty0 252--265, 2013.
\newblock \doi{10.1016/j.worlddev.2012.09.002}.

\bibitem[Grossman(1999)]{Grossman1999}
H.~Grossman.
\newblock Kleptocracy and revolutions.
\newblock \emph{Oxford Economic Papers}, 51\penalty0 (2):\penalty0 267--283,
  1999.
\newblock \doi{10.1093/oep/51.2.267}.

\bibitem[Gupta and Pelli(2021)]{Gupta2021}
R.~Gupta and M.~Pelli.
\newblock Electrification and cooking fuel choice in rural {I}ndia.
\newblock \emph{World Development}, 146, 2021.
\newblock \doi{10.1016/j.worlddev.2021.105539}.

\bibitem[Haber and Menaldo(2011)]{Haber2011}
S.~Haber and V.~Menaldo.
\newblock Do natural resources fuel authoritarianism? a reappraisal of the
  resource curse.
\newblock \emph{American Political Science Review}, 105\penalty0 (1):\penalty0
  1--26, 2011.
\newblock \doi{10.1017/S0003055410000584}.

\bibitem[Hache(2018)]{Hache2018}
E.~Hache.
\newblock Do renewable energies improve energy security in the long run?
\newblock \emph{International Economics}, 156:\penalty0 127--135, 2018.
\newblock \doi{10.1016/j.inteco.2018.01.005}.

\bibitem[Hailemariam et~al.(2021)Hailemariam, Sakutukwa, and
  Yew]{Hailemariam2021}
A.~Hailemariam, T.~Sakutukwa, and S.~Yew.
\newblock The impact of energy poverty on physical violence.
\newblock \emph{Energy Economics}, 100, 2021.
\newblock \doi{10.1016/j.eneco.2021.105336}.

\bibitem[Hanley et~al.(2007)Hanley, Shogren, and White]{Hanley2007}
N.~Hanley, J.~Shogren, and B.~White.
\newblock \emph{Environmental Economics in Theory and Practice (2nd edition)}.
\newblock Palgrave MacMillan, 2007.

\bibitem[Hanley et~al.(2013)Hanley, Shogren, and White]{Hanley2013}
N.~Hanley, J.~Shogren, and B.~White.
\newblock \emph{Introduction to Environmental Economics (2nd edition)}.
\newblock Oxford University Press, 2013.

\bibitem[Harris and Roach(2018)]{Harris2018}
J.~Harris and B.~Roach.
\newblock \emph{Environmental and Natural Resource Economics\textemdash A
  Contemporary Approach (4th edition)}.
\newblock Routledge, 2018.

\bibitem[Havranek et~al.(2016)Havranek, Horvath, and Zeynalov]{Havranek2016}
T.~Havranek, R.~Horvath, and A.~Zeynalov.
\newblock Natural resources and economic growth: A meta-analysis.
\newblock \emph{World Development}, 88:\penalty0 134--151, 2016.
\newblock \doi{10.1016/j.worlddev.2016.07.016}.

\bibitem[He(2019)]{He2019}
X.~He.
\newblock China's electrification and rural labor: Analysis with fuzzy
  regression discontinuity.
\newblock \emph{Energy Economics}, 81:\penalty0 650--660, 2019.
\newblock \doi{10.1016/j.eneco.2019.05.007}.

\bibitem[hn and Yonezawa(2021)]{Faehn2021}
T.~F. hn and H.~Yonezawa.
\newblock Emission targets and coalition options for a small, ambitious
  country: An analysis of welfare costs and distributional impacts for
  {N}orway.
\newblock \emph{Energy Economics}, 103:\penalty0 105607, 2021.
\newblock URL
  \url{https://www.sciencedirect.com/science/article/pii/S0140988321004746}.

\bibitem[Hodge(1995)]{Hodge1995}
I.~Hodge.
\newblock \emph{Environmental Economics}.
\newblock Basingstoke, 1995.

\bibitem[Hodler(2006)]{Hodler2006}
R.~Hodler.
\newblock The curse of natural resources in fractionalized countries.
\newblock \emph{European Economic Review}, 50\penalty0 (6):\penalty0
  1367--1386, 2006.
\newblock \doi{10.1016/j.euroecorev.2005.05.004}.

\bibitem[Hong et~al.(2022)Hong, Wu, and Zhang]{Hong2022}
X.~Hong, S.~Wu, and X.~Zhang.
\newblock Clean energy powers energy poverty alleviation: Evidence from
  {C}hinese micro-survey data.
\newblock \emph{Technological Forecasting and Social Change}, 182, 2022.
\newblock \doi{10.1016/j.techfore.2022.121737}.

\bibitem[Irwin et~al.(2020)Irwin, Hoxha, and Gr\'{e}pin]{Irwin2020}
B.~Irwin, K.~Hoxha, and K.~Gr\'{e}pin.
\newblock Conceptualising the effect of access to electricity on health in low-
  and middle-income countries: A systematic review.
\newblock \emph{Global Public Health}, 15\penalty0 (3):\penalty0 452--473,
  2020.
\newblock \doi{10.1080/17441692.2019.1695873}.

\bibitem[Jahangir~Alam and Kaneko(2019)]{JahangirAlam2019}
M.~Jahangir~Alam and S.~Kaneko.
\newblock The effects of electrification on school enrollment in {B}angladesh:
  Short- and long-run perspectives.
\newblock \emph{Energies}, 12\penalty0 (4), 2019.
\newblock \doi{10.3390/en12040629}.

\bibitem[Jamasb and Marantes(2012)]{Jamasb2012}
T.~Jamasb and C.~Marantes.
\newblock Electricity distribution networks: Investment and regulation, and
  uncertain demand.
\newblock In T.~Jamasb and M.~G. Pollitt, editors, \emph{The Future of
  Electricity Demand: Customers, Citizens and Loads}, pages 379--400. Cambridge
  University Press, Cambridge, 2012.
\newblock \doi{10.1017/CBO9780511996191.022}.

\bibitem[Jansen and Seebregts(2010)]{Jansen2010}
J.~C. Jansen and A.~J. Seebregts.
\newblock Long-term energy services security: What is it and how can it be
  measured and valued?
\newblock \emph{Energy Policy}, 38\penalty0 (4):\penalty0 1654--1664, 2010.
\newblock \doi{10.1016/j.enpol.2009.02.047}.

\bibitem[Jensen and Wantchekon(2004)]{Jensen2004}
N.~Jensen and L.~Wantchekon.
\newblock Resource wealth and political regimes in {A}frica.
\newblock \emph{Comparative Political Studies}, 37\penalty0 (7):\penalty0
  816--841, 2004.
\newblock \doi{10.1177/0010414004266867}.

\bibitem[Jeuland et~al.(2021)Jeuland, Fetter, Li, Pattanayak, Usmani,
  Bluffstone, Chávez, Girardeau, Hassen, Jagger, Jaime, Karumba, Köhlin,
  Lenz, Litzow, Masatsugu, Naranjo, Peters, Qin, Ruhinduka, Serrano-Medrano,
  Sievert, Sills, and Toman]{Jeuland2021}
M.~Jeuland, T.~Fetter, Y.~Li, S.~Pattanayak, F.~Usmani, R.~Bluffstone,
  C.~Chávez, H.~Girardeau, S.~Hassen, P.~Jagger, M.~Jaime, M.~Karumba,
  G.~Köhlin, L.~Lenz, E.~Litzow, L.~Masatsugu, M.~Naranjo, J.~Peters, P.~Qin,
  R.~Ruhinduka, M.~Serrano-Medrano, M.~Sievert, E.~Sills, and M.~Toman.
\newblock Is energy the golden thread? a systematic review of the impacts of
  modern and traditional energy use in low- and middle-income countries.
\newblock \emph{Renewable and Sustainable Energy Reviews}, 135, 2021.
\newblock \doi{10.1016/j.rser.2020.110406}.

\bibitem[Kahn(2020)]{Kahn2020}
M.~Kahn.
\newblock \emph{Fundamentals of Environmental and Urban Economics}.
\newblock M.E. Kahn, 2020.

\bibitem[Karpinska and \'{S}miech(2020)]{Karpinska2020}
L.~Karpinska and S.~\'{S}miech.
\newblock Conceptualising housing costs: The hidden face of energy poverty in
  {P}oland.
\newblock \emph{Energy Policy}, 147, 2020.
\newblock \doi{10.1016/j.enpol.2020.111819}.

\bibitem[Karpinska and \'{S}miech(2021)]{Karpinska2021}
L.~Karpinska and S.~\'{S}miech.
\newblock Breaking the cycle of energy poverty. will {P}oland make it?
\newblock \emph{Energy Economics}, 94, 2021.
\newblock \doi{10.1016/j.eneco.2020.105063}.

\bibitem[Kemfert(2019)]{Kemfert2019}
C.~Kemfert.
\newblock Green deal for europe: More climate protection and fewer fossil fuel
  wars.
\newblock \emph{Intereconomics}, 54\penalty0 (6):\penalty0 353--358, 2019.
\newblock \doi{10.1007/s10272-019-0853-9}.

\bibitem[Kennedy et~al.(2019)Kennedy, Mahajan, and Urpelainen]{Kennedy2019}
R.~Kennedy, A.~Mahajan, and J.~Urpelainen.
\newblock Quality of service predicts willingness to pay for household
  electricity connections in rural {I}ndia.
\newblock \emph{Energy Policy}, 129:\penalty0 319--326, 2019.
\newblock \doi{10.1016/j.enpol.2019.01.034}.

\bibitem[Keohane and Olmstead(2016)]{Keohane2016}
N.~Keohane and S.~Olmstead.
\newblock \emph{Markets and the Environment (2nd edition)}.
\newblock Washington, 2016.

\bibitem[Khandker et~al.(2012)Khandker, Barnes, and Samad]{Khandker2012}
S.~Khandker, D.~Barnes, and H.~Samad.
\newblock Are the energy poor also income poor? evidence from {I}ndia.
\newblock \emph{Energy Policy}, 47:\penalty0 1--12, 2012.
\newblock \doi{10.1016/j.enpol.2012.02.028}.

\bibitem[Khandker et~al.(2014)Khandker, Samad, Ali, and Barnes]{Khandker2014}
S.~Khandker, H.~Samad, R.~Ali, and D.~Barnes.
\newblock Who benefits most from rural electrification? evidence in {I}ndia.
\newblock \emph{Energy Journal}, 35\penalty0 (2):\penalty0 75--96, 2014.
\newblock \doi{10.5547/01956574.35.2.4}.

\bibitem[Kitchens and Fishback(2015)]{Kitchens2015}
C.~Kitchens and P.~Fishback.
\newblock Flip the switch: The impact of the {Rural Electrification
  Administration} 1935-1940.
\newblock \emph{Journal of Economic History}, 75\penalty0 (4):\penalty0
  1161--1195, 2015.
\newblock \doi{10.1017/S0022050715001540}.

\bibitem[Kleindorfer and Saad(2005)]{Kleindorfer2005}
P.~R. Kleindorfer and G.~H. Saad.
\newblock Managing disruption risks in supply chains.
\newblock \emph{Production and Operations Management}, 14\penalty0
  (1):\penalty0 53--68, 2005.
\newblock \doi{10.1111/j.1937-5956.2005.tb00009.x}.

\bibitem[Koirala and Acharya(2022)]{Koirala2022}
D.~Koirala and B.~Acharya.
\newblock Households’ fuel choices in the context of a decade-long
  load-shedding problem in {N}epal.
\newblock \emph{Energy Policy}, 162, 2022.
\newblock \doi{10.1016/j.enpol.2022.112795}.

\bibitem[Kolstad(2011)]{Kolstad2011}
C.~Kolstad.
\newblock \emph{Intermediate Environmental Economics (2nd international
  edition)}.
\newblock Oxford University Press, 2011.

\bibitem[Kolstad and S{\o}reide(2009)]{Kolstad2009}
I.~Kolstad and T.~S{\o}reide.
\newblock Corruption in natural resource management: Implications for policy
  makers.
\newblock \emph{Resources Policy}, 34\penalty0 (4):\penalty0 214--226, 2009.
\newblock \doi{10.1016/j.resourpol.2009.05.001}.

\bibitem[Koomson and Awaworyi~Churchill(2022)]{Koomson2022}
I.~Koomson and S.~Awaworyi~Churchill.
\newblock Employment precarity and energy poverty in post-apartheid {S}outh
  {A}frica: Exploring the racial and ethnic dimensions.
\newblock \emph{Energy Economics}, 110, 2022.
\newblock \doi{10.1016/j.eneco.2022.106026}.

\bibitem[Koomson et~al.(2022)Koomson, Afoakwah, and Ampofo]{Koomson2022ethnic}
I.~Koomson, C.~Afoakwah, and A.~Ampofo.
\newblock How does ethnic diversity affect energy poverty? insights from south
  africa.
\newblock \emph{Energy Economics}, 111, 2022.
\newblock \doi{10.1016/j.eneco.2022.106079}.

\bibitem[Kruyt et~al.(2009)Kruyt, van Vuuren, de~Vries, and
  Groenenberg]{Kruyt2009}
B.~Kruyt, D.~P. van Vuuren, H.~J.~M. de~Vries, and H.~Groenenberg.
\newblock Indicators for energy security.
\newblock \emph{Energy Policy}, 37\penalty0 (6):\penalty0 2166--2181, 2009.
\newblock \doi{10.1016/j.enpol.2009.02.006}.

\bibitem[Kumar and Rauniyar(2018)]{Kumar2018}
S.~Kumar and G.~Rauniyar.
\newblock The impact of rural electrification on income and education: Evidence
  from {B}hutan.
\newblock \emph{Review of Development Economics}, 22\penalty0 (3):\penalty0
  1146--1165, 2018.
\newblock \doi{10.1111/rode.12378}.

\bibitem[Landis et~al.(2021)Landis, Fredriksson, and Rausch]{Landis2021}
F.~Landis, G.~Fredriksson, and S.~Rausch.
\newblock Between- and within-country distributional impacts from harmonizing
  carbon prices in the {EU}.
\newblock \emph{Energy Economics}, 103:\penalty0 105585, 2021.
\newblock URL
  \url{https://www.sciencedirect.com/science/article/pii/S0140988321004540}.

\bibitem[Lawson(2022)]{Lawson2022}
K.~Lawson.
\newblock Electricity outages and residential fires: Evidence from {Cape Town,
  South Africa}.
\newblock \emph{South African Journal of Economics}, 2022.
\newblock \doi{10.1111/saje.12329}.

\bibitem[Le~Coq and Paltseva(2009)]{LeCoq2009}
C.~Le~Coq and E.~Paltseva.
\newblock Measuring the security of external energy supply in the european
  union.
\newblock \emph{Energy Policy}, 37\penalty0 (11):\penalty0 4474--4481, 2009.
\newblock \doi{10.1016/j.enpol.2009.05.069}.

\bibitem[Lee et~al.(2020{\natexlab{a}})Lee, Miguel, and Wolfram]{Lee2020}
K.~Lee, E.~Miguel, and C.~Wolfram.
\newblock Experimental evidence on the economics of rural electrification.
\newblock \emph{Journal of Political Economy}, 128\penalty0 (4):\penalty0
  1523--1565, 2020{\natexlab{a}}.
\newblock \doi{10.1086/705417}.

\bibitem[Lee et~al.(2020{\natexlab{b}})Lee, Miguel, and Wolfram]{Lee2020JEP}
K.~Lee, E.~Miguel, and C.~Wolfram.
\newblock Does household electrification supercharge economic development?
\newblock \emph{Journal of Economic Perspectives}, 34\penalty0 (1):\penalty0
  122--144, 2020{\natexlab{b}}.
\newblock \doi{10.1257/JEP.34.1.122}.

\bibitem[Lef\`{e}vre(2010)]{Lefevre2010}
N.~Lef\`{e}vre.
\newblock Measuring the energy security implications of fossil fuel resource
  concentration.
\newblock \emph{Energy Policy}, 38\penalty0 (4):\penalty0 1635--1644, 2010.
\newblock \doi{10.1016/j.enpol.2009.02.003}.

\bibitem[Lewis(2018)]{Lewis2018}
J.~Lewis.
\newblock Infant health, women's fertility, and rural electrification in the
  {United States}, 1930-1960.
\newblock \emph{Journal of Economic History}, 78\penalty0 (1):\penalty0
  118--154, 2018.
\newblock \doi{10.1017/S0022050718000050}.

\bibitem[Lewis and Severnini(2020)]{Lewis2020}
J.~Lewis and E.~Severnini.
\newblock Short- and long-run impacts of rural electrification: Evidence from
  the historical rollout of the {U.S.} power grid.
\newblock \emph{Journal of Development Economics}, 143, 2020.
\newblock \doi{10.1016/j.jdeveco.2019.102412}.

\bibitem[Lewis and Tietenberg(2019)]{Lewis2019}
L.~Lewis and T.~Tietenberg.
\newblock \emph{Environmental Economics and Policy (7th edition)}.
\newblock New York, 2019.

\bibitem[Lipscomb et~al.(2013)Lipscomb, Mushfiq~Mobarak, and
  Barham]{Lipscomb2013}
M.~Lipscomb, A.~Mushfiq~Mobarak, and T.~Barham.
\newblock Development effects of electrification: Evidence from the topographic
  placement of hydropower plants in {B}razil.
\newblock \emph{American Economic Journal: Applied Economics}, 5\penalty0
  (2):\penalty0 200--231, 2013.
\newblock \doi{10.1257/app.5.2.200}.

\bibitem[Litzow et~al.(2019)Litzow, Pattanayak, and Thinley]{Litzow2019}
E.~Litzow, S.~Pattanayak, and T.~Thinley.
\newblock Returns to rural electrification: Evidence from {B}hutan.
\newblock \emph{World Development}, 121:\penalty0 75--96, 2019.
\newblock \doi{10.1016/j.worlddev.2019.04.002}.

\bibitem[L\"{o}schel et~al.(2010)L\"{o}schel, Moslener, and
  R\"{u}bbelke]{Loschel2010}
A.~L\"{o}schel, U.~Moslener, and D.~T.~G. R\"{u}bbelke.
\newblock Indicators of energy security in industrialised countries.
\newblock \emph{Energy Policy}, 38\penalty0 (4):\penalty0 1665--1671, 2010.
\newblock \doi{10.1016/j.enpol.2009.03.061}.

\bibitem[Mayer et~al.(2021)Mayer, Dugan, Bachner, and Steininger]{Mayer2021}
J.~Mayer, A.~Dugan, G.~Bachner, and K.~W. Steininger.
\newblock Is carbon pricing regressive? insights from a recursive-dynamic {CGE}
  analysis with heterogeneous households for {A}ustria.
\newblock \emph{Energy Economics}, 104:\penalty0 105661, 2021.
\newblock URL
  \url{https://www.sciencedirect.com/science/article/pii/S0140988321005181}.

\bibitem[Mehlum et~al.(2006{\natexlab{a}})Mehlum, Moene, and
  Torvik]{Mehlum2006}
H.~Mehlum, K.~Moene, and R.~Torvik.
\newblock Cursed by resources or institutions?
\newblock \emph{World Economy}, 29\penalty0 (8):\penalty0 1117--1131,
  2006{\natexlab{a}}.
\newblock \doi{10.1111/j.1467-9701.2006.00808.x}.

\bibitem[Mehlum et~al.(2006{\natexlab{b}})Mehlum, Moene, and
  Torvik]{Mehlum2006EJ}
H.~Mehlum, K.~Moene, and R.~Torvik.
\newblock Institutions and the resource curse.
\newblock \emph{Economic Journal}, 116\penalty0 (508):\penalty0 1--20,
  2006{\natexlab{b}}.
\newblock \doi{10.1111/j.1468-0297.2006.01045.x}.

\bibitem[Meles(2020)]{Meles2020}
T.~Meles.
\newblock Impact of power outages on households in developing countries:
  Evidence from {E}thiopia.
\newblock \emph{Energy Economics}, 91, 2020.
\newblock \doi{10.1016/j.eneco.2020.104882}.

\bibitem[Meles et~al.(2021)Meles, Mekonnen, Beyene, Hassen, Pattanayak,
  Sebsibie, Klug, and Jeuland]{Meles2021}
T.~Meles, A.~Mekonnen, A.~Beyene, S.~Hassen, S.~Pattanayak, S.~Sebsibie,
  T.~Klug, and M.~Jeuland.
\newblock Households' valuation of power outages in major cities of {E}thiopia:
  An application of stated preference methods.
\newblock \emph{Energy Economics}, 102, 2021.
\newblock \doi{10.1016/j.eneco.2021.105527}.

\bibitem[Melnikov et~al.(2017)Melnikov, O'Neill, Dalton, and van
  Ruijven]{Melnikov2017}
N.~Melnikov, B.~O'Neill, M.~Dalton, and B.~van Ruijven.
\newblock Downscaling heterogeneous household outcomes in dynamic {CGE} models
  for energy-economic analysis.
\newblock \emph{Energy Economics}, 65:\penalty0 87--97, 2017.
\newblock \doi{10.1016/j.eneco.2017.04.023}.

\bibitem[Metcalf(2019)]{Metcalf2019}
G.~E. Metcalf.
\newblock The distributional impacts of {U.S.} energy policy.
\newblock \emph{Energy Policy}, 129:\penalty0 926--929, 2019.
\newblock \doi{10.1016/j.enpol.2019.01.076}.

\bibitem[Moniche-Bermejo(2022)]{Moniche-Bermejo2022}
A.~Moniche-Bermejo.
\newblock Do collective energy switching campaigns engage vulnerable
  households? evidence from the big switch.
\newblock \emph{Energy Policy}, 167, 2022.
\newblock \doi{10.1016/j.enpol.2022.113016}.

\bibitem[Motz(2021)]{Motz2021}
A.~Motz.
\newblock Security of supply and the energy transition: The households'
  perspective investigated through a discrete choice model with latent classes.
\newblock \emph{Energy Economics}, 97, 2021.
\newblock \doi{10.1016/j.eneco.2021.105179}.

\bibitem[Nakatani(2010)]{Nakatani2010}
K.~Nakatani.
\newblock {Restrictions on Foreign Investment in the Energy Sector for National
  Security Reasons: The Case of {J}apan}.
\newblock In A.~McHarg, editor, \emph{{Property and the Law in Energy and
  Natural Resources}}. Oxford University Press, 2010.
\newblock ISBN 9780199579853.
\newblock \doi{10.1093/acprof:oso/9780199579853.003.0016}.

\bibitem[Nawaz(2021)]{Nawaz2021}
S.~Nawaz.
\newblock Energy poverty, climate shocks, and health deprivations.
\newblock \emph{Energy Economics}, 100, 2021.
\newblock \doi{10.1016/j.eneco.2021.105338}.

\bibitem[Nie et~al.(2021)Nie, Li, and Sousa-Poza]{Nie2021}
P.~Nie, Q.~Li, and A.~Sousa-Poza.
\newblock Energy poverty and subjective well-being in {C}hina: New evidence
  from the {China Family Panel Studies}.
\newblock \emph{Energy Economics}, 103, 2021.
\newblock \doi{10.1016/j.eneco.2021.105548}.

\bibitem[Ogunro and Afolabi(2022)]{Ogunro2022}
T.~Ogunro and L.~Afolabi.
\newblock Evaluation of access to electricity and the socioeconomic effects in
  rural and urban expanses of {N}igeria.
\newblock \emph{International Journal of Social Economics}, 49\penalty0
  (1):\penalty0 124--137, 2022.
\newblock \doi{10.1108/IJSE-09-2020-0662}.

\bibitem[Ortiz et~al.(2019)Ortiz, Casquero-Modrego, and Salom]{Ortiz2019}
J.~Ortiz, N.~Casquero-Modrego, and J.~Salom.
\newblock Health and related economic effects of residential energy
  retrofitting in spain.
\newblock \emph{Energy Policy}, 130:\penalty0 375--388, 2019.
\newblock \doi{10.1016/j.enpol.2019.04.013}.

\bibitem[Oum(2019)]{Oum2019}
S.~Oum.
\newblock Energy poverty in the {Lao PDR} and its impacts on education and
  health.
\newblock \emph{Energy Policy}, 132:\penalty0 247--253, 2019.
\newblock \doi{10.1016/j.enpol.2019.05.030}.

\bibitem[Ozbafli and Jenkins(2016)]{Ozbafli2016}
A.~Ozbafli and G.~P. Jenkins.
\newblock Estimating the willingness to pay for reliable electricity supply: A
  choice experiment study.
\newblock \emph{Energy Economics}, 56:\penalty0 443--452, 2016.
\newblock \doi{10.1016/j.eneco.2016.03.025}.

\bibitem[Papyrakis and Gerlagh(2004)]{Papyrakis2004}
E.~Papyrakis and R.~Gerlagh.
\newblock The resource curse hypothesis and its transmission channels.
\newblock \emph{Journal of Comparative Economics}, 32\penalty0 (1):\penalty0
  181--193, 2004.
\newblock \doi{10.1016/j.jce.2003.11.002}.

\bibitem[Pasha et~al.(1989)Pasha, Ghaus, and Malik]{Pasha1989}
H.~A. Pasha, A.~Ghaus, and S.~Malik.
\newblock The economic cost of power outages in the industrial sector of
  {P}akistan.
\newblock \emph{Energy Economics}, 11\penalty0 (4):\penalty0 301--318, 1989.
\newblock \doi{10.1016/0140-9883(89)90046-7}.

\bibitem[Paudel(2021)]{Paudel2021}
J.~Paudel.
\newblock Why are people energy poor? evidence from ethnic fractionalization.
\newblock \emph{Energy Economics}, 102, 2021.
\newblock \doi{10.1016/j.eneco.2021.105519}.

\bibitem[Pearce and Turner(1990)]{Pearce1990}
D.~Pearce and R.~Turner.
\newblock \emph{Environmental and Natural Resource Economics}.
\newblock New York, 1990.

\bibitem[Perman et~al.(2011)Perman, Ma, Common, Maddison, and
  McGilvray]{Perman2011}
R.~Perman, Y.~Ma, M.~Common, D.~Maddison, and J.~McGilvray.
\newblock \emph{Natural Resource and Environmental Economics (4th edition)}.
\newblock Addison Wesley, 2011.

\bibitem[Peters and Sievert(2016)]{Peters2016}
J.~Peters and M.~Sievert.
\newblock Impacts of rural electrification revisited: The {A}frican context.
\newblock \emph{Revue d'Economie du Developpement}, 23\penalty0 (HS):\penalty0
  77--98, 2016.

\bibitem[Phaneuf and Requate(2017)]{Phaneuf2017}
D.~Phaneuf and T.~Requate.
\newblock \emph{A Course in Environmental Economics\textemdash Theory, Policy,
  and Practice}.
\newblock Cambridge University Press, 2017.

\bibitem[Pizer and Sexton(2019)]{Pizer2019}
W.~Pizer and S.~Sexton.
\newblock The distributional impacts of energy taxes.
\newblock \emph{Review of Environmental Economics and Policy}, 13\penalty0
  (1):\penalty0 104--123, 2019.
\newblock \doi{10.1093/reep/rey021}.

\bibitem[Poczter(2017)]{Poczter2017}
S.~Poczter.
\newblock You can't count on me: The impact of electricity unreliability on
  productivity.
\newblock \emph{Agricultural and Resource Economics Review}, 46\penalty0
  (3):\penalty0 579--602, 2017.
\newblock \doi{10.1017/age.2016.38}.

\bibitem[Podesta et~al.(2021)Podesta, Poudou, and Roland]{Podesta2021}
M.~Podesta, J.-C. Poudou, and M.~Roland.
\newblock The price impact of energy vouchers.
\newblock \emph{Energy Journal}, 42\penalty0 (3):\penalty0 27--54, 2021.
\newblock \doi{10.5547/01956574.42.3.mpod}.

\bibitem[Prakash and Munyanyi(2021)]{Prakash2021}
K.~Prakash and M.~Munyanyi.
\newblock Energy poverty and obesity.
\newblock \emph{Energy Economics}, 101, 2021.
\newblock \doi{10.1016/j.eneco.2021.105428}.

\bibitem[Purvis et~al.(2019)Purvis, Mao, and Robinson]{Purvis2019}
B.~Purvis, Y.~Mao, and D.~Robinson.
\newblock Three pillars of sustainability: in search of conceptual origins.
\newblock \emph{Sustainability Science}, 14\penalty0 (3):\penalty0 681--695,
  2019.
\newblock URL
  \url{https://link.springer.com/article/10.1007/s11625-018-0627-5}.

\bibitem[Rafi et~al.(2021)Rafi, Naseef, and Prasad]{Rafi2021}
M.~Rafi, M.~Naseef, and S.~Prasad.
\newblock Multidimensional energy poverty and human capital development:
  Empirical evidence from {I}ndia.
\newblock \emph{Energy Economics}, 101, 2021.
\newblock \doi{10.1016/j.eneco.2021.105427}.

\bibitem[Rao(2013)]{Rao2013}
N.~Rao.
\newblock Does (better) electricity supply increase household enterprise income
  in {I}ndia?
\newblock \emph{Energy Policy}, 57:\penalty0 532--541, 2013.
\newblock \doi{10.1016/j.enpol.2013.02.025}.

\bibitem[Rathi and Vermaak(2018)]{Rathi2018}
S.~Rathi and C.~Vermaak.
\newblock Rural electrification, gender and the labor market: A cross-country
  study of {India and South Africa}.
\newblock \emph{World Development}, 109:\penalty0 346--359, 2018.
\newblock \doi{10.1016/j.worlddev.2018.05.016}.

\bibitem[Rausch and Schwarz(2016)]{Rausch2016}
S.~Rausch and G.~Schwarz.
\newblock Household heterogeneity, aggregation, and the distributional impacts
  of environmental taxes.
\newblock \emph{Journal of Public Economics}, 138:\penalty0 43--57, 2016.
\newblock \doi{10.1016/j.jpubeco.2016.04.004}.

\bibitem[Rausch et~al.(2011)Rausch, Metcalf, and Reilly]{Rausch2011}
S.~Rausch, G.~E. Metcalf, and J.~M. Reilly.
\newblock Distributional impacts of carbon pricing: A general equilibrium
  approach with micro-data for households.
\newblock \emph{Energy Economics}, 33\penalty0 (SUPPL. 1):\penalty0 S20--S33,
  2011.
\newblock \doi{10.1016/j.eneco.2011.07.023}.

\bibitem[Reichl et~al.(2013)Reichl, Schmidthaler, and Schneider]{Reichl2013}
J.~Reichl, M.~Schmidthaler, and F.~Schneider.
\newblock The value of supply security: The costs of power outages to austrian
  households, firms and the public sector.
\newblock \emph{Energy Economics}, 36:\penalty0 256--261, 2013.
\newblock \doi{10.1016/j.eneco.2012.08.044}.

\bibitem[Riahi et~al.(2022)Riahi, Schaeffer, Arango, Calvin, Guivarch,
  Hasegawa, Jiang, Kriegler, Matthews, Peters, Rao, Robertson, Sebbit,
  Steinberger, Tavoni, and van Vuuren]{Riahi2022IPCC}
K.~Riahi, R.~Schaeffer, J.~Arango, K.~Calvin, C.~Guivarch, T.~Hasegawa,
  K.~Jiang, E.~Kriegler, R.~Matthews, G.~Peters, A.~Rao, S.~Robertson, A.~M.
  Sebbit, J.~Steinberger, M.~Tavoni, and D.~van Vuuren.
\newblock Mitigation pathways compatible with long-term goals.
\newblock In P.~R. Shukla, J.~Skea, R.~Slade, A.~A. Khourdajie, R.~van Diemen,
  D.~McCollum, M.~Pathak, S.~Some, P.~Vyas, R.~Fradera, M.~Belkacemi,
  A.~Hasija, G.~Lisboa, S.~Luz, and J.~Malley, editors, \emph{Climate Change
  2022: Mitigation of Climate Change\textemdash Contribution of Working Group
  III to the Sixth Assessment Report of the Intergovernmental Panel on Climate
  Change}. Cambridge University Press, Cambridge, 2022.

\bibitem[Richmond and Urpelainen(2019)]{Richmond2019}
J.~Richmond and J.~Urpelainen.
\newblock Electrification and appliance ownership over time: Evidence from
  rural {I}ndia.
\newblock \emph{Energy Policy}, 133, 2019.
\newblock \doi{10.1016/j.enpol.2019.06.070}.

\bibitem[Robinson et~al.(2006)Robinson, Torvik, and Verdier]{Robinson2006}
J.~Robinson, R.~Torvik, and T.~Verdier.
\newblock Political foundations of the resource curse.
\newblock \emph{Journal of Development Economics}, 79\penalty0 (2):\penalty0
  447--468, 2006.
\newblock \doi{10.1016/j.jdeveco.2006.01.008}.

\bibitem[Rosas-Flores et~al.(2017)Rosas-Flores, Bakhat, Rosas-Flores, and
  Fern\'{a}ndez~Zayas]{RosasFlores2017}
J.~Rosas-Flores, M.~Bakhat, D.~Rosas-Flores, and J.~Fern\'{a}ndez~Zayas.
\newblock Distributional effects of subsidy removal and implementation of
  carbon taxes in {M}exican households.
\newblock \emph{Energy Economics}, 61:\penalty0 21--28, 2017.
\newblock \doi{10.1016/j.eneco.2016.10.021}.

\bibitem[Ross(1999)]{Ross1999}
M.~Ross.
\newblock The political economy of the resource curse.
\newblock \emph{World Politics}, 51\penalty0 (2):\penalty0 297--322, 1999.
\newblock \doi{10.1017/S0043887100008200}.

\bibitem[Sachs and Warner(2001)]{Sachs2001}
J.~Sachs and A.~Warner.
\newblock The curse of natural resources.
\newblock \emph{European Economic Review}, 45\penalty0 (4-6):\penalty0
  827--838, 2001.
\newblock \doi{10.1016/S0014-2921(01)00125-8}.

\bibitem[Sadath and Acharya(2017)]{Sadath2017}
A.~Sadath and R.~Acharya.
\newblock Assessing the extent and intensity of energy poverty using
  {Multidimensional Energy Poverty Index}: Empirical evidence from households
  in {I}ndia.
\newblock \emph{Energy Policy}, 102:\penalty0 540--550, 2017.
\newblock \doi{10.1016/j.enpol.2016.12.056}.

\bibitem[Saelim(2019)]{Saelim2019}
S.~Saelim.
\newblock Carbon tax incidence on household consumption: Heterogeneity across
  socio-economic factors in {T}hailand.
\newblock \emph{Economic Analysis and Policy}, 62:\penalty0 159--174, 2019.
\newblock \doi{10.1016/j.eap.2019.02.003}.

\bibitem[Saing(2018)]{Saing2018}
C.~Saing.
\newblock Rural electrification in {C}ambodia: does it improve the welfare of
  households?
\newblock \emph{Oxford Development Studies}, 46\penalty0 (2):\penalty0
  147--163, 2018.
\newblock \doi{10.1080/13600818.2017.1340443}.

\bibitem[Salmon and Tanguy(2016)]{Salmon2016}
C.~Salmon and J.~Tanguy.
\newblock Rural electrification and household labor supply: Evidence from
  {N}igeria.
\newblock \emph{World Development}, 82:\penalty0 48--68, 2016.
\newblock \doi{10.1016/j.worlddev.2016.01.016}.

\bibitem[Sambodo and Novandra(2019)]{Sambodo2019}
M.~Sambodo and R.~Novandra.
\newblock The state of energy poverty in {I}ndonesia and its impact on welfare.
\newblock \emph{Energy Policy}, 132:\penalty0 113--121, 2019.
\newblock \doi{10.1016/j.enpol.2019.05.029}.

\bibitem[Samuelson(2014)]{Samuelson2014}
R.~D. Samuelson.
\newblock The unexpected challenges of using energy intensity as a policy
  objective: Examining the debate over the {APEC} energy intensity goal.
\newblock \emph{Energy Policy}, 64:\penalty0 373--381, 2014.
\newblock \doi{10.1016/j.enpol.2013.09.020}.

\bibitem[Sanghvi(1982)]{Sanghvi1982}
A.~Sanghvi.
\newblock Economic costs of electricity supply interruptions. us and foreign
  experience.
\newblock \emph{Energy Economics}, 4\penalty0 (3):\penalty0 180--198, 1982.
\newblock \doi{10.1016/0140-9883(82)90017-2}.

\bibitem[Schmidthaler et~al.(2015)Schmidthaler, Cohen, Reichl, and
  Schmidinger]{Schmidthaler2015}
M.~Schmidthaler, J.~Cohen, J.~Reichl, and S.~Schmidinger.
\newblock The effects of network regulation on electricity supply security: a
  european analysis.
\newblock \emph{Journal of Regulatory Economics}, 48\penalty0 (3):\penalty0
  285--316, 2015.
\newblock \doi{10.1007/s11149-015-9277-z}.

\bibitem[Sedai et~al.(2021{\natexlab{a}})Sedai, Jamasb, Nepal, and
  Miller]{Sedai2021EE}
A.~Sedai, T.~Jamasb, R.~Nepal, and R.~Miller.
\newblock Electrification and welfare for the marginalized: Evidence from
  india.
\newblock \emph{Energy Economics}, 102, 2021{\natexlab{a}}.
\newblock \doi{10.1016/j.eneco.2021.105473}.

\bibitem[Sedai et~al.(2021{\natexlab{b}})Sedai, Vasudevan, Pena, and
  Miller]{Sedai2021}
A.~Sedai, R.~Vasudevan, A.~Pena, and R.~Miller.
\newblock Does reliable electrification reduce gender differences? evidence
  from {I}ndia.
\newblock \emph{Journal of Economic Behavior and Organization}, 185:\penalty0
  580--601, 2021{\natexlab{b}}.
\newblock \doi{10.1016/j.jebo.2021.03.015}.

\bibitem[Sedai et~al.(2022)Sedai, Nepal, and Jamasb]{Sedai2022}
A.~Sedai, R.~Nepal, and T.~Jamasb.
\newblock Electrification and socio-economic empowerment of women in {I}ndia.
\newblock \emph{Energy Journal}, 43\penalty0 (2):\penalty0 215--238, 2022.
\newblock \doi{10.5547/01956574.43.2.ased}.

\bibitem[Serra and Fierro(1997)]{Serra1997}
P.~Serra and G.~Fierro.
\newblock Outage costs in {C}hilean industry.
\newblock \emph{Energy Economics}, 19\penalty0 (4):\penalty0 417--434, 1997.
\newblock \doi{10.1016/S0140-9883(97)01017-7}.

\bibitem[Shi et~al.(2022)Shi, Xu, Gao, Zhang, and Chang]{Shi2022}
H.~Shi, H.~Xu, W.~Gao, J.~Zhang, and M.~Chang.
\newblock The impact of energy poverty on agricultural productivity: The case
  of china.
\newblock \emph{Energy Policy}, 167, 2022.
\newblock \doi{10.1016/j.enpol.2022.113020}.

\bibitem[Sievert and Steinbuks(2020)]{Sievert2020}
M.~Sievert and J.~Steinbuks.
\newblock Willingness to pay for electricity access in extreme poverty:
  Evidence from sub-{S}aharan {A}frica.
\newblock \emph{World Development}, 128, 2020.
\newblock \doi{10.1016/j.worlddev.2019.104859}.

\bibitem[Singh and Inglesi-Lotz(2020)]{Singh2020}
K.~Singh and R.~Inglesi-Lotz.
\newblock The role of energy poverty on economic growth in sub-{S}aharan
  {A}frican countries.
\newblock \emph{Economics of Energy and Environmental Policy}, 10\penalty0 (1),
  2020.
\newblock \doi{10.5547/2160-5890.9.2.KSIN}.

\bibitem[Smith(2015)]{Smith2015}
B.~Smith.
\newblock The resource curse exorcised: Evidence from a panel of countries.
\newblock \emph{Journal of Development Economics}, 116:\penalty0 57--73, 2015.
\newblock \doi{10.1016/j.jdeveco.2015.04.001}.

\bibitem[Sovacool and Mukherjee(2011)]{Sovacool2011}
B.~Sovacool and I.~Mukherjee.
\newblock Conceptualizing and measuring energy security: A synthesized
  approach.
\newblock \emph{Energy}, 36\penalty0 (8):\penalty0 5343--5355, 2011.
\newblock \doi{10.1016/j.energy.2011.06.043}.

\bibitem[Sovacool(2013)]{Sovacool2013}
B.~K. Sovacool.
\newblock An international assessment of energy security performance.
\newblock \emph{Ecological Economics}, 88:\penalty0 148--158, 2013.
\newblock \doi{10.1016/j.ecolecon.2013.01.019}.

\bibitem[Steinbuks and Foster(2010)]{Steinbuks2010}
J.~Steinbuks and V.~Foster.
\newblock When do firms generate? evidence on in-house electricity supply in
  africa.
\newblock \emph{Energy Economics}, 32\penalty0 (3):\penalty0 505--514, 2010.

\bibitem[Stirling(2010)]{Stirling2010}
A.~Stirling.
\newblock Multicriteria diversity analysis. a novel heuristic framework for
  appraising energy portfolios.
\newblock \emph{Energy Policy}, 38\penalty0 (4):\penalty0 1622--1634, 2010.
\newblock \doi{10.1016/j.enpol.2009.02.023}.

\bibitem[Suehiro(2008)]{Suehiro2008}
S.~Suehiro.
\newblock Energy intensity as an index of energy conservation: problems in
  international comparison of energy intensity of gdp and estimate using
  sector-based approach.
\newblock \emph{IEEJ Energy Journal}, 7:\penalty0 58--71, 2008.

\bibitem[Tagliapietra et~al.(2020)Tagliapietra, Occhiali, Nano, and
  Kalcik]{Tagliapietra2020}
S.~Tagliapietra, G.~Occhiali, E.~Nano, and R.~Kalcik.
\newblock The impact of electrification on labour market outcomes in nigeria.
\newblock \emph{Economia Politica}, 37\penalty0 (3):\penalty0 737--779, 2020.
\newblock \doi{10.1007/s40888-020-00189-2}.

\bibitem[{te Brake}(1975)]{teBrake1975}
W.~H. {te Brake}.
\newblock Air pollution and fuel crises in preindustrial {L}ondon, 1250-1650.
\newblock \emph{Technology and Culture}, 16\penalty0 (3):\penalty0 337--359,
  1975.
\newblock URL \url{http://www.jstor.org/stable/3103030}.

\bibitem[Teller-Elsberg et~al.(2016)Teller-Elsberg, Sovacool, Smith, and
  Laine]{Teller-Elsberg2016}
J.~Teller-Elsberg, B.~Sovacool, T.~Smith, and E.~Laine.
\newblock Fuel poverty, excess winter deaths, and energy costs in vermont:
  Burdensome for whom?
\newblock \emph{Energy Policy}, 90:\penalty0 81--91, 2016.
\newblock \doi{10.1016/j.enpol.2015.12.009}.

\bibitem[Thomas and Urpelainen(2018)]{Thomas2018}
D.~Thomas and J.~Urpelainen.
\newblock Early electrification and the quality of service: Evidence from rural
  {I}ndia.
\newblock \emph{Energy for Sustainable Development}, 44:\penalty0 11--20, 2018.
\newblock \doi{10.1016/j.esd.2018.02.004}.

\bibitem[Thomas et~al.(2020)Thomas, Harish, Kennedy, and
  Urpelainen]{Thomas2020}
D.~Thomas, S.~Harish, R.~Kennedy, and J.~Urpelainen.
\newblock The effects of rural electrification in {I}ndia: An instrumental
  variable approach at the household level.
\newblock \emph{Journal of Development Economics}, 146, 2020.
\newblock \doi{10.1016/j.jdeveco.2020.102520}.

\bibitem[Tietenberg and Lewis(2018)]{Tietenberg2018}
T.~Tietenberg and L.~Lewis.
\newblock \emph{Environmental and Natural Resource Economics (11th edition)}.
\newblock New York, 2018.

\bibitem[Tinbergen(1952)]{Tinbergen1952}
J.~Tinbergen.
\newblock \emph{On the Theory of Economic Policy}.
\newblock North Holland, Amsterdam, 1952.

\bibitem[Tishler(1993)]{Tishler1993}
A.~Tishler.
\newblock Optimal production with uncertain interruptions in the supply of
  electricity. estimation of electricity outage costs.
\newblock \emph{European Economic Review}, 37\penalty0 (6):\penalty0
  1259--1274, 1993.
\newblock \doi{10.1016/0014-2921(93)90134-V}.

\bibitem[Tong et~al.(2019)Tong, Zhang, Zheng, Caldeira, Shearer, Hong, Qin, and
  Davis]{Tong2019}
D.~Tong, Q.~Zhang, Y.~Zheng, K.~Caldeira, C.~Shearer, C.~Hong, Y.~Qin, and
  S.~Davis.
\newblock Committed emissions from existing energy infrastructure jeopardize
  1.5 °c climate target.
\newblock \emph{Nature}, 572\penalty0 (7769):\penalty0 373--377, 2019.
\newblock \doi{10.1038/s41586-019-1364-3}.

\bibitem[Torvik(2009)]{Torvik2009}
R.~Torvik.
\newblock Why do some resource-abundant countries succeed while others do not?
\newblock \emph{Oxford Review of Economic Policy}, 25\penalty0 (2):\penalty0
  241--256, 2009.
\newblock \doi{10.1093/oxrep/grp015}.

\bibitem[Toto(2022)]{Toto2022}
R.~V. Toto.
\newblock Residential welfare-loss from electricity supply interruptions in
  {S}outh {A}frica: Cost-benefit analysis of distributed energy resource
  subsidy programs.
\newblock \emph{Economics of Energy and Environmental Policy}, 11\penalty0
  (1):\penalty0 51--77, 2022.
\newblock \doi{10.5547/2160-5890.11.1.RTOT}.

\bibitem[Tovar Rea\~{n}os and W\"{o}lfing(2018)]{TovarReanos2018}
M.~Tovar Rea\~{n}os and N.~W\"{o}lfing.
\newblock Household energy prices and inequality: Evidence from {G}erman
  microdata based on the {EASI} demand system.
\newblock \emph{Energy Economics}, 70:\penalty0 84--97, 2018.
\newblock \doi{10.1016/j.eneco.2017.12.002}.

\bibitem[Turner et~al.(1994)Turner, Pearce, and Bateman]{Turner1994}
R.~Turner, D.~Pearce, and I.~Bateman.
\newblock \emph{Environmental Economics\textemdash An Elementary Introduction}.
\newblock Harlow, 1994.

\bibitem[van~de Walle et~al.(2017)van~de Walle, Ravallion, Mendiratta, and
  Koolwal]{vandeWalle2017}
D.~van~de Walle, M.~Ravallion, V.~Mendiratta, and G.~Koolwal.
\newblock Long-term gains from electrification in rural {I}ndia.
\newblock \emph{World Bank Economic Review}, 31\penalty0 (2):\penalty0
  385--411, 2017.
\newblock \doi{10.1093/wber/lhv057}.

\bibitem[Van Der~Ploeg(2011)]{VanDerPloeg2011}
F.~Van Der~Ploeg.
\newblock Natural resources: Curse or blessing?
\newblock \emph{Journal of Economic Literature}, 49\penalty0 (2):\penalty0
  366--420, 2011.
\newblock \doi{10.1257/jel.49.2.366}.

\bibitem[van~der Ploeg and Poelhekke(2009)]{vanderPloeg2009}
F.~van~der Ploeg and S.~Poelhekke.
\newblock Volatility and the natural resource curse.
\newblock \emph{Oxford Economic Papers}, 61\penalty0 (4):\penalty0 727--760,
  2009.

\bibitem[van~der Ploeg and Poelhekke(2010)]{vanderPloeg2010}
F.~van~der Ploeg and S.~Poelhekke.
\newblock The pungent smell of ``red herrings'': Subsoil assets, rents,
  volatility and the resource curse.
\newblock \emph{Journal of Environmental Economics and Management}, 60\penalty0
  (1):\penalty0 44--55, 2010.
\newblock \doi{10.1016/j.jeem.2010.03.003}.

\bibitem[Vandyck et~al.(2021)Vandyck, Weitzel, Wojtowicz, {Rey Los Santos},
  Maftei, and Riscado]{Vandyck2021}
T.~Vandyck, M.~Weitzel, K.~Wojtowicz, L.~{Rey Los Santos}, A.~Maftei, and
  S.~Riscado.
\newblock Climate policy design, competitiveness and income distribution: A
  macro-micro assessment for 11 {EU} countries.
\newblock \emph{Energy Economics}, 103:\penalty0 105538, 2021.
\newblock URL
  \url{https://www.sciencedirect.com/science/article/pii/S0140988321004151}.

\bibitem[Vicente(2010)]{Vicente2010}
P.~Vicente.
\newblock Does oil corrupt? evidence from a natural experiment in {West
  Africa}.
\newblock \emph{Journal of Development Economics}, 92\penalty0 (1):\penalty0
  28--38, 2010.
\newblock \doi{10.1016/j.jdeveco.2009.01.005}.

\bibitem[Wagner et~al.(2021)Wagner, Rieger, Bedi, Vermeulen, and
  Demena]{Wagner2021}
N.~Wagner, M.~Rieger, A.~Bedi, J.~Vermeulen, and B.~Demena.
\newblock The impact of off-grid solar home systems in {K}enya on energy
  consumption and expenditures.
\newblock \emph{Energy Economics}, 99, 2021.
\newblock \doi{10.1016/j.eneco.2021.105314}.

\bibitem[Williams(2011)]{Williams2011}
A.~Williams.
\newblock Shining a light on the resource curse: An empirical analysis of the
  relationship between natural resources, transparency, and economic growth.
\newblock \emph{World Development}, 39\penalty0 (4):\penalty0 490--505, 2011.
\newblock \doi{10.1016/j.worlddev.2010.08.015}.

\bibitem[Wills(1997)]{Wills1997}
I.~Wills.
\newblock \emph{Economics and the Environment\textemdash A Signalling and
  Incentives Approach}.
\newblock Allen \& Unwin, 1997.

\bibitem[Wirawan and Gultom(2021)]{Wirawan2021}
H.~Wirawan and Y.~Gultom.
\newblock The effects of renewable energy-based village grid electrification on
  poverty reduction in remote areas: The case of {I}ndonesia.
\newblock \emph{Energy for Sustainable Development}, 62:\penalty0 186--194,
  2021.
\newblock \doi{10.1016/j.esd.2021.04.006}.

\bibitem[Woo et~al.(2021)Woo, Tishler, Zarnikau, and Chen]{Woo2021}
C.~K. Woo, A.~Tishler, J.~Zarnikau, and Y.~Chen.
\newblock Average residential outage cost estimates for the lower 48 states in
  the {US}.
\newblock \emph{Energy Economics}, 98, 2021.
\newblock \doi{10.1016/j.eneco.2021.105270}.

\bibitem[Wu et~al.(2022)Wu, Zhang, and Qian]{Wu2022}
L.~Wu, S.~Zhang, and H.~Qian.
\newblock Distributional effects of {C}hina's national emissions trading scheme
  with an emphasis on sectoral coverage and revenue recycling.
\newblock \emph{Energy Economics}, 105:\penalty0 105770, 2022.
\newblock URL
  \url{https://www.sciencedirect.com/science/article/pii/S0140988321006125}.

\bibitem[Wu et~al.(2021)Wu, Zeng, Gong, and Chen]{Wu2021}
W.-P. Wu, W.-K. Zeng, S.-W. Gong, and Z.-G. Chen.
\newblock Does energy poverty reduce rural labor wages? evidence from
  {C}hina’s rural household survey.
\newblock \emph{Frontiers in Energy Research}, 9, 2021.
\newblock \doi{10.3389/fenrg.2021.670026}.

\bibitem[Xu and Chen(2019)]{Xu2019}
X.~Xu and C.-F. Chen.
\newblock Energy efficiency and energy justice for {U.S.} low-income
  households: An analysis of multifaceted challenges and potential.
\newblock \emph{Energy Policy}, 128:\penalty0 763--774, 2019.
\newblock \doi{10.1016/j.enpol.2019.01.020}.

\bibitem[Zhang et~al.(2019)Zhang, Shi, Zhang, and Xiao]{Zhang2019}
T.~Zhang, X.~Shi, D.~Zhang, and J.~Xiao.
\newblock Socio-economic development and electricity access in developing
  economies: A long-run model averaging approach.
\newblock \emph{Energy Policy}, 132:\penalty0 223--231, 2019.
\newblock \doi{10.1016/j.enpol.2019.05.031}.

\bibitem[Zhang et~al.(2021)Zhang, Shu, Yi, and Wang]{Zhang2021}
Z.~Zhang, H.~Shu, H.~Yi, and X.~Wang.
\newblock Household multidimensional energy poverty and its impacts on physical
  and mental health.
\newblock \emph{Energy Policy}, 156, 2021.
\newblock \doi{10.1016/j.enpol.2021.112381}.

\end{thebibliography}

\end{document}